\documentclass{aa}

\usepackage{txfonts}
\usepackage{color, colortbl}
\usepackage[colorlinks, citecolor=blue, linkcolor=blue, urlcolor=blue]{hyperref}

\usepackage{graphicx}
\usepackage{txfonts}
\usepackage{pdflscape}	
\usepackage{color, colortbl}

\usepackage{newtxtext,newtxmath}

\usepackage[T1]{fontenc}
\usepackage{ae,aecompl}

\usepackage{pdflscape}	

\usepackage{listings}
\usepackage{pdflscape}
\usepackage{longtable}
\usepackage{wasysym}

\usepackage{graphicx}	
\usepackage{amsmath}	
\usepackage{amssymb}	
\usepackage{multirow}

\usepackage{soul}
\usepackage{blindtext}

\begin{document}

\title{The Galactic Bulge exploration I.: The period-absolute\,magnitude-metallicity relations for RR~Lyrae stars for $G_{\rm BP}$, $V$, $G$, $G_{\rm RP}$, $I$, $J$, $H$, and $K_{\rm s}$ passbands using $Gaia$ DR3 parallaxes}

\author{Z.~Prudil\inst{1,2}, A.~Kunder\inst{3}, I. D\'ek\'any\inst{1}, A.~J~Koch-Hansen\inst{1} }

\institute{
Astronomisches Rechen-Institut, Zentrum f{\"u}r Astronomie der Universit{\"a}t Heidelberg, M{\"o}nchhofstr. 12-14, D-69120 Heidelberg, Germany  
\and European Southern Observatory, Karl-Schwarzschild-Strasse 2, 85748 Garching bei M\"{u}nchen, Germany; \email{Zdenek.Prudil@eso.org} 
\and Saint Martin's University, 5000 Abbey Way SE, Lacey, WA, 98503}

\date{\today}

\abstract
{We present a new set of period–absolute magnitude–metallicity (PMZ) relations for single-mode RR Lyrae stars calibrated for the optical $G_{\rm BP}$, $V$, $G$, $G_{\rm RP}$, near-infrared $I$, $J$, $H$, and $K_{\rm s}$ passbands. We compiled a large dataset (over $100$ objects) of fundamental and first-overtone RR~Lyrae pulsators consisting of mean intensity magnitudes, reddenings, pulsations properties, iron abundances, and parallaxes measured by the \textit{Gaia} astrometric satellite in its third data release. Our newly calibrated PMZ relations encapsulate the most up-to-date ingredients in terms of both data and methodology. They are aimed to be used in conjunction with large photometric surveys targeting the Galactic bulge, including the Optical Gravitational Lensing Experiment (OGLE), the Vista Variables in the Vía Láctea Survey (VVV), and the \textit{Gaia} catalog. In addition, our Bayesian probabilistic approach provides accurate uncertainty estimates of the predicted absolute magnitudes of individual RR Lyrae stars. Our derived PMZ relations provide consistent results when compared to benchmark distances to Globular clusters NGC\,6121 (also known as M4), NGC\,5139 (also known as omega Cen), and Large and Small Magellanic Clouds, which are stellar systems rich in RR~Lyrae stars. Lastly, our $K_{\rm s}$-band PMZ relations match well with the previously published PMZ relations based on Gaia data and accurately predict the distance toward the prototype of this class of variables, the eponymic RR~Lyr itself.}

\keywords{Stars: variables: RR~Lyrae -- methods: statistical – methods: data analysis -- parallaxes}
\titlerunning{RR~Lyrae PMZ relations}
\authorrunning{Prudil et al.}
\maketitle

\section{Introduction} \label{sec:Intro}

RR~Lyrae stars are old, helium-burning, mostly radially pulsating horizontal branch giants. They are divided into three main groups based on their pulsation mode; fundamental mode (RRab), first-overtone (RRc), and double-mode pulsators (RRd) pulsating both in fundamental and first-overtone mode. The pulsation properties of all three subclasses have been studied extensively \citep[e.g.,][]{Bono1996,Szabo2010,Netzel2015-02,Smolec2015-BL,Skarka2020,Molnar2022RRlyr,Netzel2022}. The fundamental mode RR~Lyrae pulsators often serve as distance and metallicity indicators toward old stellar systems within our Galaxy \citep[e.g.,][]{Braga2015,Skowron2016,Dekany2021} and beyond \citep[e.g.,][]{Sarajedini2006,Sarajedini2009,Fiorentino2012M32,Savino2022}.

In particular, the connection between pulsation periods, absolute magnitudes, and metallicities (PMZ relations) of RR Lyrae stars allows us to infer their distances from photometric time series, and it has been one of their most practical features. RR~Lyrae PMZ relations have granted opportunities to study structures of old stellar systems \citep[e.g.,][]{Dekany2013,MartinezVazquez2016,Jacyszyn-Dobrzeniecka2017,Jacyszyn-Dobrzeniecka2020} and their distances \citep[e.g.,][]{Braga2015,MartinezVazquez2019,Bhardwaj2020,Fabrizio2021,Bhardwaj2021}. There are both theoretical and empirical approaches to calibrate the PMZ relations in the literature for various passbands ranging from optical to far-infrared \citep[e.g.,][]{Bono2003,Catelan2004Temp,Neeley2015,Marconi2015,Neeley2017,Marconi2018PLZ,Muraveva2018,Neeley2019PLZ,Garofalo2022}. 

Empirical studies in the past relied on large samples of RR~Lyrae stars ($\approx$400) with metallicities on the \citet{Zinn1984} metallicity scale \citep[combining both spectroscopically and photometrically estimated metallicities, e.g.,][]{Dambis2013,Muraveva2018,Muhie2021}. The distances in the aforementioned studies were either based on statistical parallax analysis \citep[e.g.,][]{Dambis2009,Dambis2013} or on trigonometric parallaxes measured by the \textit{Gaia} space mission \citep[e.g.,][]{Muraveva2018,Neeley2019PLZ,Layden2019,Muhie2021,Garofalo2022} or lastly using distances to RR~Lyrae rich globular clusters \citep[e.g.,][]{Neeley2015,Bhardwaj2021}. The photometric information often came from various sources, mainly large photometric surveys covering a broad range of passbands.

In this work, we aim to empirically calibrate the PMZ relations for three major photometric surveys targeting (among others) dense stellar regions in the Milky Way; the Optical Gravitational Lensing Experiment \citep[OGLE,][]{Udalski2015}, the Vista Variables in the V\'ia L\'actea survey \citep[VVV,][]{Minniti2010}, and $Gaia$ astrometric mission \citep{Gaia2016}. The combination of these surveys provides optical (OGLE, $Gaia$) and near-infrared (VVV) photometry for tens of thousands of RR~Lyrae stars in crowded stellar regions in the Milky Way. The precise calibration of PMZ relations will allow for accurate distance determination and possible avenues to treat extinction directly based on the color excess of RR~Lyrae pulsators. This will enable us to probe the structure of the Galactic bulge from the point of view of old population pulsators \citep[similarly to studies like][]{Dekany2013,Pietrukowicz2015,Prudil2019OOspat,Du2020,Molnar2022} and its kinematical properties \citep{Kunder2016,Prudil2019Kin,Kunder2020,Du2020}.

Previous studies have used both theoretical and empirical PMZ relations to determine distances toward the RR~Lyrae population in the Galactic bulge. These relations were not directly calibrated to the photometric systems of OGLE and VVV. They required re-calibration mainly since the OGLE $V$-band is partially different from the standard Johnson $V$-passband \citep[see][]{Udalski2015}, and also the VVV $JHK_{\rm s}$ passbands differ from those in the \textit{Two-Micron Sky Survey} \citep[2MASS,][]{Cutri2003,Skrutskie2006}. In addition, the metallicity term in the aforementioned PMZ relations was usually calibrated to a different metallicity scale than the estimated metallicities for the bulge RR~Lyrae population, thus requiring further conversion \citep[e.g., for $Gaia$ photometry in a study by][]{Muraveva2018}. Furthermore, previous studies of the Galactic bulge relied mostly on external extinction maps and did not use RR~Lyrae stars themself as tracers of extinction. This motivated the decision to provide PMZ relations that are more appropriate for RR~Lyrae stars toward the Galactic bulge. Our subsequent papers will use this calibration to investigate the structure and kinematics of the bulge RR~Lyrae population. For the newly derived PMZ relations, we used purely geometrically determined distances by the $Gaia$ astrometric mission \citep{Lindegren2021,GaiaDR32022}, combined with publicly available data on RR~Lyrae stars in the Solar neighborhood. In particular, we aimed to use a homogenous metallicity scale among calibrating stars that will allow direct use of photometric metallicities derived from OGLE $I$-band photometry for RR~Lyrae variables toward the Galactic bulge.

This paper is structured as follows: In Section~\ref{sec:PMZcalib}, we describe our assembled dataset and analysis of the data. The subsequent Section~\ref{sec:CalibPMZ} outlines our statistical approach to the PMZ calibration and derived equations together with their covariance matrices. In Sect.~\ref{sec:TestingComparPMZ}, we test our newly derived relation with the literature values for some of the RR~Lyrae-rich stellar systems. Finally, Section~\ref{sec:Summary} summarizes our results.

\section{Dataset for calibration of the PMZ relations} \label{sec:PMZcalib}

We based our calibration sample on stars with iron abundances collected in \citet{Crestani2021Alpha} and \citet{Dekany2021}. Similar to the calibration by \citet{Dekany2021}, individual spectroscopic measurements of [\ion{Fe}{i}/H] and [\ion{Fe}{ii}/H] were collected from the literature and shifted to the common metallicity scale defined by \citet[][abbreviated as CFCS]{For2011chem,Chadid2017,Sneden2017,Crestani2021} using offsets from \citet{Dekany2021}. Each star in our calibration sample had multiple published iron abundance measurements and an associated uncertainty. We obtained the pulsation properties (pulsation period, $P$) for individual variables in the spectroscopic dataset from various sources, mainly from the International Variable Star Index \citep[VSX,][]{Watson2006VSX}.

\begin{figure*}
\includegraphics[width=2\columnwidth]{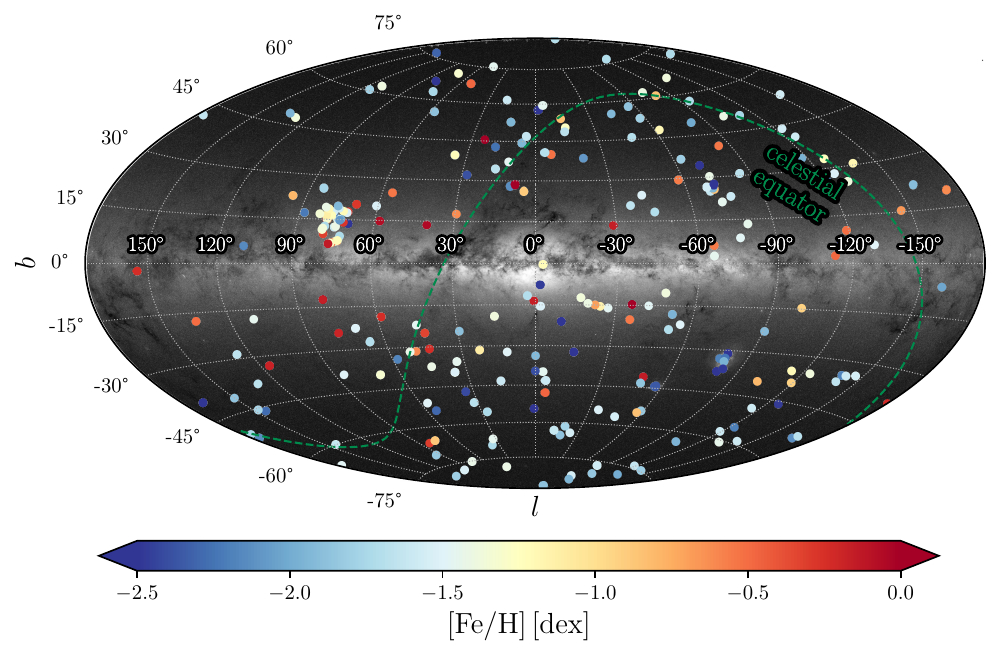}
\caption{Spatial distribution in the Galactic coordinates of our RR~Lyrae calibration dataset with color-coding representing objects metallicity. $Gaia$'s all-sky star density map is underpinned in the background. The overdensity at $l \approx 80$\,deg and $b \approx 10$\,deg is due to RR~Lyrae stars observed by the $Kepler$ space telescope and followed up spectroscopically by \citet{Nemec2013}. \textit{Image credit: Gaia Data Processing and Analysis Consortium (DPAC); A. Moitinho / A. F. Silva / M. Barros / C. Barata, University of Lisbon, Portugal; H. Savietto, Fork Research, Portugal.}}
\label{fig:SampleMap}
\end{figure*}

Often pulsation periods are quoted without their appropriate errors; thus, to account for possible minor uncertainties in $P$ we used the following equation to approximate the uncertainties of pulsation periods $\sigma_{P}$:
\begin{equation} \label{eq:ErrPerEqu}
\sigma_{P} = 1.001 \cdot P - 0.999 \cdot P \\.
\end{equation}
This approach fits $\sigma_{P}$ over an approximate baseline of $1000$\,days. For a general RRab variable with a pulsation period equal to $0.5$\,day Eq.~\ref{eq:ErrPerEqu} yields $\sigma_{P} = 0.001$\,day. In addition, we converted pulsation periods of the first-overtone (FO) pulsators into fundamental mode using the following equation from \citet{Iben1971} and \citet{Braga2016}: 
\begin{equation} \label{eq:RRC_fundam}
\text{log}_{10}(P) = \text{log}_{10} (P_{\rm FO}) + 0.127 \\.
\end{equation}
For the entire spectroscopic dataset, values for parallaxes $\varpi$, and their uncertainties $\sigma_{\varpi}$, were obtained from the third data release (DR3) of the $Gaia$ mission \citep[and subsequently corrected for the zero point offset \footnote{Using a code \url{https://gitlab.com/icc-ub/public/gaiadr3_zeropoint}.}, by][]{Lindegren2021}. To account for interstellar reddening and to ensure its homogenous treatment, we used 2D extinction maps from \citet{Schlegel1998} to obtain $E(B-V)$ color excesses and its uncertainties toward individual stars and re-calibrated reddening laws from \citet{Schlafly2011}. The spatial distribution of our spectroscopic dataset in the Galactic coordinates is depicted in Fig.~\ref{fig:SampleMap}.

\subsection{Mean intensity magnitudes}

The intensity magnitudes were obtained from various sources and subsequently transformed into the OGLE and VVV passbands. We emphasize that our RR Lyrae sample sizes vary between the different passbands, and we list the final number of RR~Lyrae stars for each calibration after applying quality cuts (see Section~\ref{sec:CalibPMZ} and Eq.~\ref{eq:QualityCut}).

In the case of the $G_{\rm BP}$, $G$ and $G_{\rm RP}$-bands photometry, we utilized the $Gaia$ RR~Lyrae catalog with their associated photometric and pulsation properties \citep{Clementini2022}. As the mean intensity magnitudes, $m_{G_{\rm BP}}$, $m_{G}$, $m_{G_{\rm RP}}$, we used intensity-averaged magnitudes in the aforementioned bands together with their RR~Lyrae subclassification, pulsation periods, and uncertainties on pulsation periods. In total, we used $240$ ($201$ RRab and $39$ RRc stars) RR~Lyrae variables for $G_{\rm BP}$, $G$ and $G_{\rm RP}$-bands, respectively.

For the $V$-passband, we obtained photometry for RR~Lyrae stars and subsequently their mean intensity magnitudes, $m_{V}$, from the All-Sky Automated Survey for Supernovae \citep[ASAS-SN,][]{Shappee2014,Jayasinghe2018}. This sample contained $166$ RR~Lyrae stars ($136$ RRab and $30$ RRc pulsators). The same match was conducted for the $I$-band photometry from \citet{Dekany2021}, which yielded $128$ RR~Lyrae stars in total ($104$ RRab and $24$ RRc pulsators) with mean intensity magnitudes $m_{I}$. We note that $I$-band photometry for the calibration RR~Lyrae stars came from the predecessor of ASAS-SN, the All Sky Automated Survey \citep[ASAS,][]{Pojmanski1997,Szczygiel2009} and was collected from work by \citet{Dekany2021}.

The mean intensity magnitudes for $V$ and $I$-passbands were obtained through Fourier decomposition of photometric light curves using an approach described in \citet{Petersen1986}. We optimized the following Fourier light curve decomposition:
\begin{equation} \label{eq:FourSer}
m\left ( t \right ) = m + \sum_{k=1}^{n} A_{k} \cdot \text{cos} \left (2\pi k \vartheta + \varphi_{k} \right ) \\.
\end{equation}
In Equation~\ref{eq:FourSer}, $m$, represents the mean intensity magnitude, $A_{k}$ and $\varphi_{k}$ stand for amplitudes and phases. The $n$ denotes the degree of the fit which we adapted for each light curve using the same approach as in \citet{Prudil2019OOspat}. The $\vartheta$ represents the phase function defined as:
\begin{equation} \label{eq:phasing}
\vartheta = \left(\text{HJD}-M_{0}\right)/P \\,
\end{equation}
where HJD is a Heliocentric Julian Date representing the time of the observation, and $M_{0}$ stands for the time of brightness maximum\footnote{It is important to note, that the $M_{0}$ was estimated for each survey separately.}. 

The $V$ and $I$-passbands from the ASAS-SN and ASAS surveys were directly transformed into the OGLE photometric system \citep{Udalski2015} using common stars between surveys. We used the following linear relations to transform intensity magnitudes:
\begin{gather} \label{eq:Conver-mags}
m_{V}^{\rm OGLE} = 1.003 \cdot m_{V} + 0.006 \\
m_{I}^{\rm OGLE} = 0.992 \cdot m_{I} + 0.107 \hspace{1.75cm}.
\end{gather}
In the case of the $J$ and $H$-bands, we used the photometry provided by the 2MASS and estimated mean intensity magnitudes, $m_{J}$ and $m_{H}$, from \citet{Braga2019}. Since 2MASS provides, in most cases, single epoch observations at different pulsation phases, we needed to correct for the luminosity variation during the pulsation cycle to obtain mean intensity magnitudes. For this purpose, we utilized near-infrared photometric templates derived by \citet{Braga2019} in combination with optical, $V$-band, photometric data from ASAS \citep[to accurately estimate $M_{0}$][]{Pojmanski1997}. The choice of using the ASAS data stemmed from the similar time baseline between ASAS and 2MASS, which allowed better constraints to the 2MASS single epoch observations (pulsation period, time of brightness maxima, amplitude) necessary for template fitting and obtaining the $m_{J}$ and $m_{H}$. In the end, we had $115$ RR~Lyrae stars ($97$ RRab and $18$ RRc variables) with mean intensity magnitudes $m_{J}$ and $111$ RR~Lyrae stars ($94$ RRab and $17$ RRc variables) with mean intensity magnitudes $m_{H}$.

The mean intensity magnitudes in $K_{\rm s}$-band were acquired from several sources, particularly from studies by \citet{Layden2019} and \citet{Braga2019}. We also used the 2MASS survey, where the same approach as in the $J$ and $H$-passbands was used to correct for the luminosity variations as a function of the pulsation cycle. A combination of the studies and surveys mentioned earlier resulted in $155$ RR~Lyrae stars ($129$ RRab and $26$ RRc variables) with mean intensity magnitudes $m_{K_{\rm s}}$. The dataset provided by \citet{Layden2019} offered the mean $K_{\rm s}$-band magnitudes transformed into the 2MASS photometric system, while \citet{Braga2019} photometry was also calibrated to the 2MASS photometric system. Therefore, acquired $J$, $H$, and $K_{\rm s}$ mean intensity magnitudes needed to be transformed into the VVV photometric system. We used the equations from the CASU website\footnote{\url{http://casu.ast.cam.ac.uk/surveys-projects/vista/technical/photometric-properties/sky-brightness-variation/view}} and color terms $J-K_{\rm s}$ for individual stars to convert our $m_{J}$, $m_{H}$, and $m_{K_{\rm s}}$ magnitudes into the VVV photometric system. 

\section{Calibration of the PMZ} \label{sec:CalibPMZ}

In this work, we aim to improve the following relation:
\begin{equation} \label{eq:PMZ_to_calib}
M = \alpha\,\text{log}_{10}(P) + \beta\,\text{[Fe/H]} + \gamma \\,
\end{equation}
where $M$ is the absolute magnitude in the given passband (in our study $G_{\rm BP}$, $V$, $G$, $G_{\rm RP}$, $I$, $J$, $H$, and $K_{\rm s}$). The pulsation periods and metallicities are denoted as $P$ and [Fe/H], together with parameters of the PMZ marked as $\alpha$, $\beta$, and $\gamma$. Although our dataset is smaller in comparison to previous studies \citep[e.g.,][]{Dambis2013,Muraveva2018,Muhie2021}, our calibration is based on a homogeneous metalicity scale that is the same as what is used for the Galactic bulge RR~Lyrae stars. This allows for more accurate reddening and distance to be derived, especially in the $G_{\rm BP}$, $V$, $G$, $G_{\rm RP}$, and $I$-passbands which are more affected by both reddening and the effect of [Fe/H] metallicity on absolute magnitude. It is important to emphasize that we used both pulsation periods and metallicities to estimate absolute magnitudes for optical ($G_{\rm BP}$ and $V$) passbands instead of only metallicities as was done in previous studies \citep[e.g.,][]{Catelan2004,Muraveva2018}. We used both based on a mild anticorrelation (Pearson correlation coefficient equal to $-0.47$) between pulsation periods and absolute magnitudes (calculated using parallaxes), for both passbands. In addition, the combination of both parameters will decrease uncertainty in the distance, particularly in cases where the metallicity of a given RR~Lyrae variable is unknown or highly uncertain.

Our dataset for each passband $\mathbf{D}$ consists of vectors $\mathbf{d}^{k}$ that contain the following information for individual stars $k$:
\begin{equation}
\mathbf{d}^{k} = \left\lbrace \text{log}_{\rm 10}(P), \text{[Fe/H]}, \varpi, m, E(B-V)  \right\rbrace ^{k} \\.
\end{equation}
To properly account for errors in our catalog and accurately estimate the PMZ parameters, we employed the following approach utilizing the Bayesian framework where the posterior probability $p(\boldsymbol{\theta}\,|\,\mathbf{D})$ is equal to:
\begin{equation} \label{eq:BayPos}
p(\boldsymbol{\theta}\,|\,\mathbf{D}) \propto p(\mathbf{D}\,|\,\boldsymbol{\theta}) \, p(\boldsymbol{\theta}) \\.
\end{equation}
The $\theta$ represents the model parameters $\boldsymbol{\theta}_{i} = \left\lbrace \alpha_{i}, \beta_{i}, \gamma_{i}, \varepsilon_{M_{i}} \right\rbrace $ of the PMZ relation (see equation~\ref{eq:PMZ_to_calib}), $\varepsilon_{M_{i}}$ represents the intrinsic scatter in the PMZ relation, and the $p(\boldsymbol{\theta})$ is the prior probability of the parameter set $\boldsymbol{\theta}$. In our approach, we assumed the prior probability for the intrinsic scatter in the PMZ, $\varepsilon_{M}$, that was in the form of a Jeffreys log-uniform prior \citep{Jaynes1968}:
\begin{equation} \label{eq:BayPrior}
p(\varepsilon_{M_{i}}) = 1 / \varepsilon_{M_{i}} \\.
\end{equation}
For the $\alpha$, $\beta$, and $\gamma$ parameters we selected mildly informative uniform priors, $\mathcal{U}$, based on values from \citet[][see their Table~6 for $I$, $J$, $H$, and $K_{\rm s}$]{Marconi2015}. We used the following priors for both the optical $G_{\rm BP}$, $V$, $G$, and $G_{\rm RP}$ bands:
\begin{gather}
p(\alpha_{i}, \beta_{i}, \gamma_{i}) = \mathcal{U}(-2.5 < \alpha < 0.5)\\
\hspace{2.1cm} \mathcal{U}(0.0 < \beta < 0.5)\\
\hspace{2.1cm} \mathcal{U}(-1.5 < \gamma < 1.5) \hspace{1cm}.
\end{gather}
The marginalized likelihood $p(\mathbf{D}\,|\,\boldsymbol{\theta})$ for $k$ number of stars can be written as such:
\begin{equation} \label{eq:BayLikel}
p(\mathbf{D}\,|\,\boldsymbol{\theta}_{i}) = \prod_{k=1}^{K} p(\mathbf{d}\,|\,\boldsymbol{\theta}) = \prod_{k=1}^{K} \mathcal{N}(M_{\rm data}\,|\,M_{\rm model}, \sigma_{M_{\rm tot}}) \\,
\end{equation}
The $M_{\rm data}$ represents absolute magnitudes estimated using collected photometric and astrometric data. The $M_{\rm model}$ stands for absolute magnitudes estimated through Eq.~\ref{eq:PMZ_to_calib}. The $\mathcal{N}$ represents the normal distribution with a mean $\mu$ and standard deviation $\sigma$ for a variable $x$, $N (x \,|\, \mu, \sigma^2)$. In the equations above, $R_{i}$ denotes the extinction coefficient in a given passband, where we selected a baseline from \citet{Cardelli1989} with $R_{V} = 3.1$, for individually calibrated passband, we used extinction coefficients from \citet[][for $V$, $I$, and $K_{\rm s}$]{Schlafly2011}, and for the $Gaia$ $G_{\rm BP}$, $G$, and $G_{\rm RP}$-passbands we used the color-dependent extinction coefficients from \citet[][see their Eq.~1 and Table~1 for details]{Hertzsprung2018ExtRed}. The total uncertainty, $\sigma_{M_{\rm tot}}$, encompassed the individual uncertainties for $M_{\rm data}$, $M_{\rm model}$ and $\varepsilon_{M_{i}}$ added in quadrature.

We implemented a set of selection criteria for all involved passbands to ensure we use high-quality data. The general requirements on RR~Lyrae stars used for PMZ calibration were that re-normalized unit weight error (RUWE\footnote{The RUWE parameter estimates the quality of the $Gaia$ astrometric solution.}) was lower than $1.4$, uncertainty on mean intensity magnitude was $\sigma_{m_{i}} < 0.1$\,mag, and we used only stars outside the highly reddened regions of the MW using condition $\vert b \vert > 15$\,deg. The first and third general conditions prevented the possible problems with astrometric and photometric solutions (e.g., blending). The second condition was primarily applied in order to uncertain mean intensity magnitudes in $J$. In addition, we implemented two conditions on parallax significance and reddening:
\begin{equation} \label{eq:QualityCut}
\varpi / \sigma_{\varpi} > 20 \hspace{0.4cm} \text{and} \hspace{0.4cm} E(B-V) < 0.3 \\ .
\end{equation}
These conditions ensured that we used highly precise parallaxes with stars only weakly affected by reddening. The latter was particularly important for optical passbands where reddening and selecting a reddening law could play a significant role in PMZ calibration.

\begin{figure}
\includegraphics[width=\columnwidth]{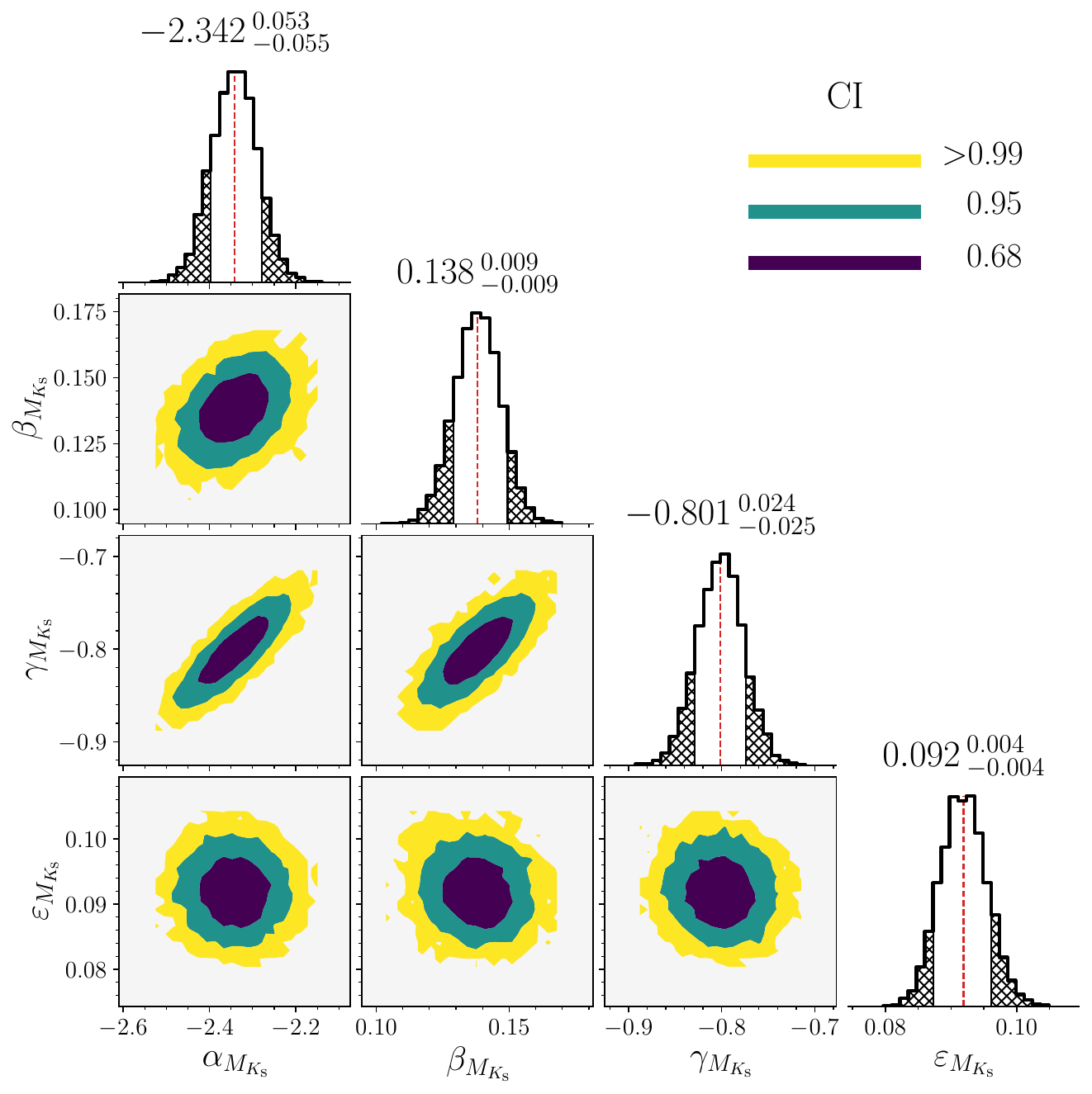}
\caption{Posterior probability distributions of the parameters of the PMZ relation for the $K_{\rm s}$-passband.}
\label{fig:CornerKs}
\end{figure}

Using the aforementioned approach and selection criteria, we proceeded to estimate the parameters of the individual PMZ relations. To explore the parameter space, $\boldsymbol{\theta}$, we employed the Markov Chain Monte Carlo Ensemble sampler implemented in the \textsc{emcee} package \citet{Foreman-Mackey2013} to maximize the posterior probability defined in the Eq.~\ref{eq:BayPos}. We ran \textsc{emcee} with $100$ walkers for $5000$ steps. Using \textsc{emcee} we thinned the chains by $\tau=10$ and selected the first $4000$ steps as a burn-in, which resulted in $10000$ samples for the posterior distributions. The distribution of $\boldsymbol{\theta}$ parameters for $K_{\rm s}$, $H$, $J$, $I$, $G_{\rm RP}$, $G$,  $G_{\rm BP}$, and $V$-bands is shown in Figures \ref{fig:CornerKs}, \ref{fig:CornerH}, \ref{fig:CornerJ}, \ref{fig:CornerI}, \ref{fig:CornerGrp}, \ref{fig:CornerG}, \ref{fig:CornerGbp}, and \ref{fig:CornerV}, respectively. The parameters of PMZ relations for eight passbands, together with the total number of variables, $N$, used in calibration, can be found in the following equations: 
\begin{align}
\begin{split}\label{eq:1}
M_{K_{\rm s}} = {}& -2.342\,\text{log}_{10}(P) + 0.138\,\text{[Fe/H]} - 0.801 , \hspace{0.05cm} N = 97
\end{split}\\
\begin{split}\label{eq:2} 
M_{H} = {}& -2.250\,\text{log}_{10}(P) + 0.157\,\text{[Fe/H]} - 0.665 , \hspace{0.05cm} N = 72
\end{split}\\
\begin{split}\label{eq:3}
M_{J} = {}& -1.799\,\text{log}_{10}(P) + 0.160\,\text{[Fe/H]} - 0.378 , \hspace{0.05cm} N = 64
\end{split}\\
\begin{split}\label{eq:3}
M_{I} = {}& -1.292\,\text{log}_{10}(P) + 0.196\,\text{[Fe/H]} + 0.197 , \hspace{0.05cm} N = 79
\end{split}\\
\begin{split}\label{eq:4} 
M_{G_{\rm RP}} = {}& -1.464\,\text{log}_{10}(P) + 0.167\,\text{[Fe/H]} + 0.113 , \hspace{0.05cm} N = 110
\end{split}\\
\begin{split}\label{eq:5} 
M_{G} = {}& -0.950\,\text{log}_{10}(P) + 0.202\,\text{[Fe/H]} + 0.614 , \hspace{0.05cm} N = 110
\end{split}\\
\begin{split}\label{eq:6}
M_{V} = {}& -0.582\,\text{log}_{10}(P) + 0.224\,\text{[Fe/H]} + 0.890 , \hspace{0.05cm} N = 112
\end{split}\\
\begin{split}\label{eq:7}
M_{G_{\rm BP}} = {}& -0.593\,\text{log}_{10}(P) + 0.228\,\text{[Fe/H]} + 0.913 , \hspace{0.05cm} N = 107
\end{split}
\end{align}
Their covariance matrices with associated intrinsic scatter are the following:
\begin{equation}
\text{Cov}_{K_{\rm s}} = 
\begin{bmatrix}
\sigma_{\alpha_{K_{\rm s}}} & \sigma_{\beta_{K_{\rm s}}} & \sigma_{\gamma_{K_{\rm s}}} \\
\hline
0.00287 & 0.00017 & 0.00113 \\
0.00017 & 0.00008 & 0.00017\\
0.00113 & 0.00017 & 0.00062 \\
\end{bmatrix} \\ \hspace{0.3cm} \varepsilon_{M_{K_{\rm s}}} = 0.092
\end{equation}
\begin{equation}
\text{Cov}_{H} = 
\begin{bmatrix}
\sigma_{\alpha_{H}} & \sigma_{\beta_{H}} & \sigma_{\gamma_{H}} \\
\hline
0.01007 & 0.00052 & 0.00375 \\
0.00052 & 0.00030 & 0.00062 \\
0.00375 & 0.00062 & 0.00216 \\
\end{bmatrix} \\ \hspace{0.3cm} \varepsilon_{M_{H}} = 0.151
\end{equation}
\begin{equation}
\text{Cov}_{J} = 
\begin{bmatrix}
\sigma_{\alpha_{J}} & \sigma_{\beta_{J}} & \sigma_{\gamma_{J}} \\
\hline
0.01397 & 0.00049 & 0.00476 \\
0.00049 & 0.00041 & 0.00081 \\
0.00476 & 0.00081 & 0.00275 \\
\end{bmatrix} \\ \hspace{0.3cm} \varepsilon_{M_{J}} = 0.171
\end{equation}
\begin{equation}
\text{Cov}_{I} = 
\begin{bmatrix}
\sigma_{\alpha_{I}} & \sigma_{\beta_{I}} & \sigma_{\gamma_{I}} \\
\hline
0.00496 & 0.00027 & 0.00200 \\
0.00027 & 0.00016 & 0.00033 \\
0.00200 & 0.00033 & 0.00118 \\
\end{bmatrix} \\ \hspace{0.3cm} \varepsilon_{M_{I}} = 0.112
\end{equation}
\begin{equation}
\text{Cov}_{G_{\rm RP}} = 
\begin{bmatrix}
\sigma_{\alpha_{G_{\rm RP}}} & \sigma_{\beta_{G_{\rm RP}}} & \sigma_{\gamma_{G_{\rm RP}}} \\
\hline
0.00348 & 0.00019 & 0.00135 \\
0.00019 & 0.00009 & 0.00019 \\
0.00135 & 0.00019 & 0.00072 \\
\end{bmatrix} \\ \hspace{0.3cm} \varepsilon_{M_{G_{\rm RP}}} = 0.107
\end{equation}
\begin{equation}
\text{Cov}_{G} = 
\begin{bmatrix}
\sigma_{\alpha_{G}} & \sigma_{\beta_{G}} & \sigma_{\gamma_{G}} \\
\hline
0.00432 & 0.00023 & 0.00166 \\
0.00023 & 0.00011 & 0.00024 \\
0.00166 & 0.00024 & 0.00089 \\
\end{bmatrix} \\ \hspace{0.3cm} \varepsilon_{M_{G}} = 0.118
\end{equation}
\begin{equation}
\text{Cov}_{V} = 
\begin{bmatrix}
\sigma_{\alpha_{V}} & \sigma_{\beta_{V}} & \sigma_{\gamma_{V}} \\
\hline
0.00564 & 0.00032 & 0.00220 \\
0.00032 & 0.00015 & 0.00032 \\
0.00220 & 0.00032 & 0.00119 \\
\end{bmatrix} \\ \hspace{0.3cm} \varepsilon_{M_{V}} = 0.140 \hspace{0.1cm} .
\end{equation}
\begin{equation}
\text{Cov}_{G_{\rm BP}} = 
\begin{bmatrix}
\sigma_{\alpha_{G_{\rm BP}}} & \sigma_{\beta_{G_{\rm BP}}} & \sigma_{\gamma_{G_{\rm BP}}} \\
\hline
0.00498 & 0.00027 & 0.00192 \\
0.00027 & 0.00013 & 0.00028 \\
0.00192 & 0.00028 & 0.00104 \\
\end{bmatrix} \\ \hspace{0.3cm} \varepsilon_{M_{G_{\rm BP}}} = 0.126
\end{equation}
In the derived PMZ relations, we see a linear dependence of individual coefficients and passbands. As we move from near-infrared toward the optical passband, the value of the $\alpha$ coefficient decreases \citep[consequently the importance of pulsation period,][]{Longmore1986,Dallora2004}, and the value of $\beta$ increases (the relevance of metallicity). These effects are well known, particularly on the theoretical side, where the near-infrared $K_{\rm s}$ is shown to have less dependence on metallicity and also less dependence on evolutionary effects \citep[see, e.g.,][]{Bono2003,Catelan2004,Marconi2015,Marconi2018PLZ}. Observations also indicate that there is a decrease in intrinsic scatter when finding $M_{K_{\rm s}}$ \citep[e.g.,][]{Neeley2019PLZ,Layden2019,Cusano2021}.

\begin{figure}
\includegraphics[width=\columnwidth]{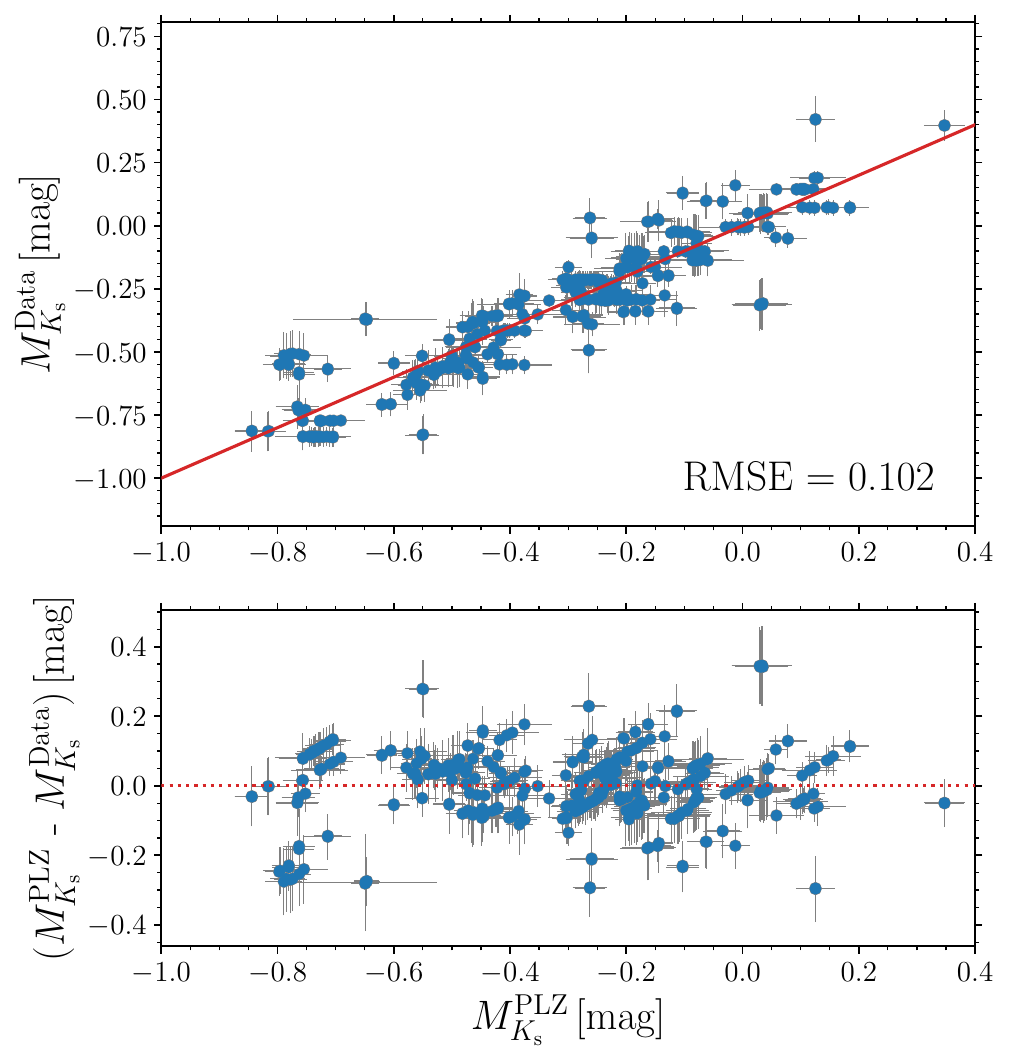}
\caption{Comparison of the absolute magnitudes predicted by our $K_{\rm s}$-band PMZ relation with absolute magnitudes calculated based on $Gaia$ data.}
\label{fig:PMZ_comK}
\end{figure}

Together with the assumed intrinsic scatter in the PMZ, $\varepsilon_{M}$ the absolute magnitude uncertainty calculation for a given RR~Lyrae with a parameter vector $\mathbf{v} = \left\lbrace \text{log}_{\rm 10}(P), \text{[Fe/H]}, 1.0 \right\rbrace $ is in the following form:
\begin{equation}
\begin{split} 
\sigma_{M}^{2} = & \mathbf{v} \times \text{Cov} \times \mathbf{v}^{\rm T} + \left[\sigma_{P} \cdot \alpha / \left(P\,\text{log}\left(10\right)\right) \right]^{2} \\ & + \left( \beta \cdot \sigma_{\text{[Fe/H]}} \right)^{2} + \varepsilon_{M}^{2} \hspace{0.5cm}.
\end{split}
\end{equation}
Comparison between absolute magnitude predictions based on our new PMZ relations and estimated absolute magnitudes calculated using $Gaia$ parallaxes and dereddened intensity mean magnitudes are shown in Figures~\ref{fig:PMZ_comK}, \ref{fig:PMZ_comJ}, \ref{fig:PMZ_comI}, \ref{fig:PMZ_comG}, and \ref{fig:PMZ_comV}.

\section{Testing predicted absolute magnitudes and comparison of PMZ relations} \label{sec:TestingComparPMZ}

In the following section, we examined the derived PMZ relations using the publicly available data on stellar systems with precisely derived distances using methods other than RR~Lyrae PMZ relations. Also, we compare the derived relation for PMZ in $K_{\rm s}$-band with those derived in the literature.

\subsection{On the predicted absolute magnitudes} \label{subsec:TestingPMZ}

We selected only systems with a significant number of RR~Lyrae stars and metallicities for testing predicted absolute magnitudes on the same metallicity scale as in our PMZ calibration. We picked globular clusters NGC\,6121 and NGC\,5139 ($\equiv$ M\,4 and $\omega$\,Centauri, respectively) that have distances and their uncertainties derived using the $Gaia$ parallaxes \citep[][$\varpi^{\rm {NGC\,6121}} = 0.556 \pm 0.010$\,mas and $\varpi^{\rm {NGC\,5139}} = 0.193 \pm 0.009$\,mas, respectively]{Vasiliev2021}. These distances are in agreement within two sigmas with the estimates based on RR~Lyrae pulsators \citep{Neeley2015,Braga2018Ome}. We used available mean intensity magnitudes in optical and near-infrared passbands from \citet{Stetson2014,Braga2016,Braga2018Ome} and their listed spectroscopic metallicities and pulsation periods when available. In the case of NGC\,6121, we used a value for metallicity from \citet{Carretta2009} catalog since it is very close to the CFCS scale \citep[difference below $-0.1$\,dex,][]{Crestani2021}. To account for extinction, we used reddening maps and reddening laws from \citet{Schlegel1998} and \citet{Schlafly2011}, respectively.

In the case of the Large and Small Magellanic Clouds (from hereon LMC and SMC), we used $Gaia$ and OGLE-IV photometry ($G_{\rm BP}$, $V$ and $I$-bands), and stellar classifications from OGLE-IV \citep{Soszynski2016LMCSMC}. We re-analyzed the OGLE photometry using the Eq.~\ref{eq:FourSer} and estimated photometric metallicities using relations from \citet{Dekany2021}, which are on the same scale as our PMZ calibration. For the near-infrared data, we obtained $K_{\rm s}$ mean intensity magnitudes for $8599$ LMC RR~Lyrae stars \citep{Cusano2021} and $473$ SMC RR~Lyrae pulsators from \citet{Muraveva2018SMC}. For the LMC and SMC distances, we used estimates from \citet[][]{Pietrzynski2019} and \citet{Graczyk2020SMC} based on late-type eclipsing binary stars. For dereddening the obtained mean intensity magnitudes, we used reddening maps from \citet{Schlegel1998} and reddening laws from \citet{Schlafly2011}. 

\begin{figure*}
\includegraphics[width=2\columnwidth]{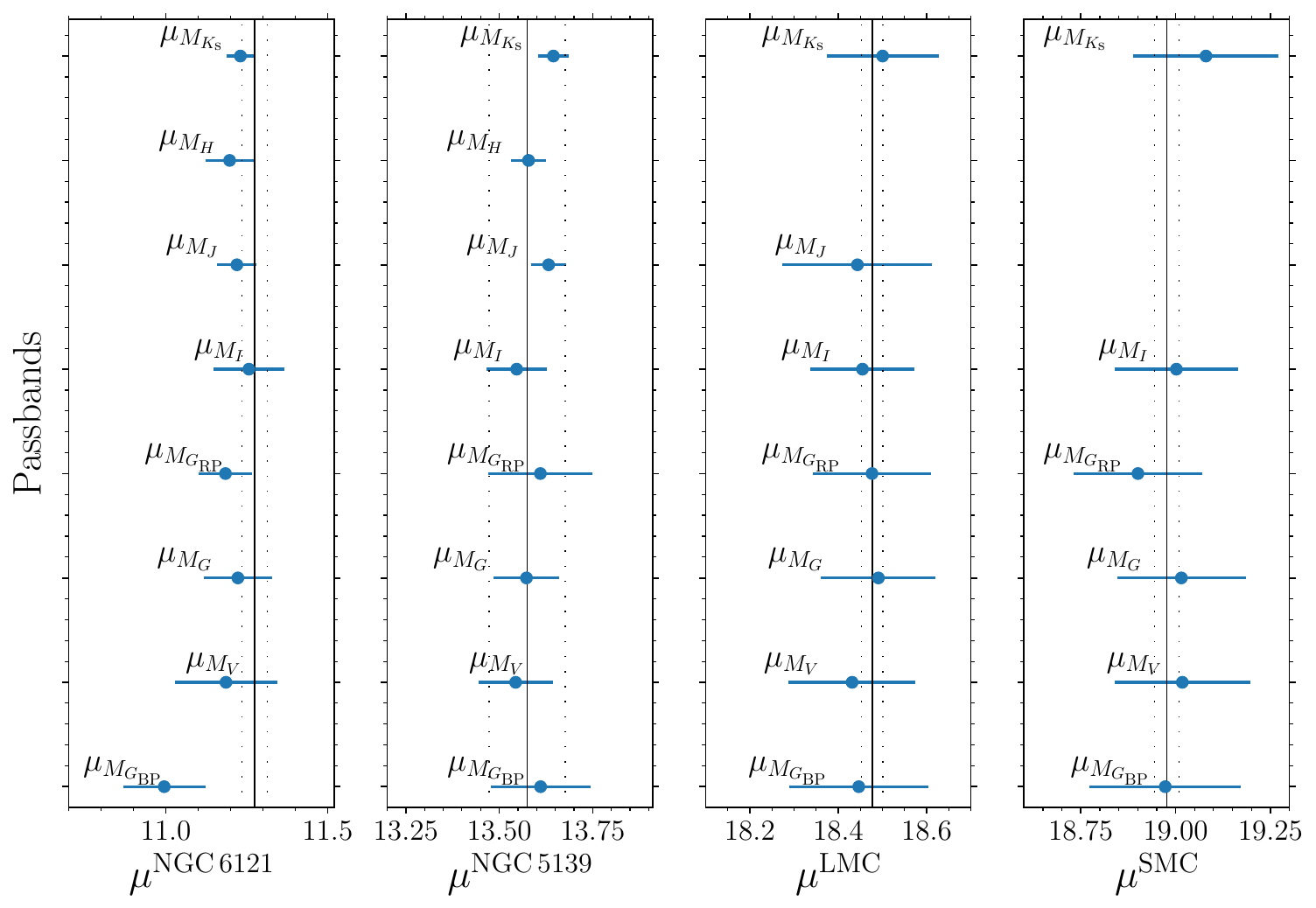}
\caption{Comparison of distance moduli for four systems (NGC\,6121, NGC\,5139, LMC, and SMC) with precise distances and uncertainties derived using different methods \citep[other than RR~Lyrae PMZ relations,][]{Pietrzynski2019,Graczyk2020SMC,Vasiliev2021} . From the left, the plots show a comparison for globular clusters NGC\,6121, NGC\,5139, LMC, and SMC, using distances and their uncertainties depicted with solid and dotted black lines in all three plots.}
\label{fig:PMZ_comparisonSys}
\end{figure*}

We calculated the distance moduli of individual stars and estimated the peak of the distance modulus distribution for each system and passband using the Kernel density estimate routine \citep[KDE, implemented in the \texttt{scipy} library,][]{scipy}. The error is estimated as the dispersion of the distance modulus distribution. It is worth noting that this method may not be ideal for the SMC and LMC, as they possess a significant depth extent that can lead to an increased level of measured distance dispersion
\citep[e.g.,][]{Jacyszyn-Dobrzeniecka2017, Tatton2021}. In Figure~\ref{fig:PMZ_comparisonSys}, we compare our PMZ relations with distances to known stellar systems rich in RR~Lyrae with sufficient data coverage. In general, the calculated distances using the absolute magnitude relations presented here mostly agree within one $\sigma$ of the literature distance when using the optical $G_{\rm BP}$, $V$, $G$, $G_{\rm RP}$, $I$, and near-infrared $J$, $H$, and $K_{\rm s}$ passbands. The determined distance moduli are listed in Table~\ref{tab:DistanceModuli} together with their literature values. In the case of the optical $G_{\rm BP}$ and $G$-bands, we see a significant offset (around $0.2 - 0.25$\,mag) in distance modulus from the literature value for NGC\,6121. This discrepancy is most likely caused by a significant differential reddening toward NGC\,6121 (on average $E(B-V) = 0.5$\,mag), the actual proximity of NGC\,6121 (2D reddening maps provide only projected reddening), and in part, color-dependent extinction coefficients for $G_{\rm BP}$ and $G$-bands. The offset mostly disappears when we use 3D extinction maps \citep{Green2018dust,Green2019}. In addition, the offset in the aforementioned optical bands is only observed for NGC\,6121. In the other three comparisons, we see a good agreement with the literature distance moduli.

In conclusion, our optical and near-infrared PMZ relations agree with literature values for NGC\,6121, NGC\,5139, LMC, and SMC distances. We detect an offset from the canonical values that various effects, e.g., reddening, can cause for the $G_{\rm BP}$ and $G$-passbands. 

\begin{table}
\setlength{\tabcolsep}{3.1pt}
\setlength\extrarowheight{2pt}
\caption{List of distance moduli for tested stellar systems using derived PMZ relations. The first column contains the source for calculated distance moduli (passbands and literature value). The second and third columns represent the distance moduli of two globular clusters, NGC\,6121 and NGC\,5139. The last two columns list distance moduli for Magellanic Clouds.}
\footnotesize{
\begin{tabular}{lcccc}\hline 
       Band      & $\mu^{\rm NGC\,6121}$ & $\mu^{\rm NGC\,5139}$ & $\mu^{\rm LMC}$  & $\mu^{\rm SMC}$  \\ 
 & [mag]      & [mag]      & [mag] & [mag] \\ \hline 
$G_{\rm BP}$ & $11.00~\pm~0.13$      & $13.61~\pm~0.13$      & $18.45~\pm~0.16$ & $18.97~\pm~0.20$ \\
$V$          & $11.25~\pm~0.16$      & $13.54~\pm~0.10$      & $18.43~\pm~0.14$ & $19.02~\pm~0.18$ \\
$G$          & $11.12~\pm~0.11$      & $13.57~\pm~0.09$      & $18.49~\pm~0.13$ & $19.02~\pm~0.17$ \\
$G_{\rm RP}$ & $11.17~\pm~0.08$      & $13.61~\pm~0.14$      & $18.48~\pm~0.13$ & $18.90~\pm~0.17$ \\
$I$          & $11.26~\pm~0.11$      & $13.55~\pm~0.08$      & $18.45~\pm~0.12$ & $18.91~\pm~0.16$ \\
$J$          & $11.24~\pm~0.06$      & $13.63~\pm~0.05$      & $18.44~\pm~0.17$ & --               \\
$H$          & $11.16~\pm~0.07$      & $13.58~\pm~0.05$      & --               & --               \\
$K_{\rm s}$  & $11.27~\pm~0.04$      & $13.65~\pm~0.04$      & $18.50~\pm~0.13$ & $19.08~\pm~0.19$ \\ \hline
 Lit. & $11.28~\pm~0.04$ & $13.58~\pm~0.10$ & $18.48~\pm~0.02$ & $18.98~\pm~0.03$ \\ \hline 
\end{tabular}}
\label{tab:DistanceModuli}
\end{table}

\subsection{Comparison with earlier studies} \label{subsec:ComparsionPrevStudies}

We compare our derived PMZ relation in the $K_{\rm s}$-band with relations derived in different studies \citep[e.g.,][]{Bono2003,Sollima2006,Sollima2008,Borissova2009,Dambis2013,Muraveva2015,Bhardwaj2021,Muhie2021,Bhardwaj2023}. We focused only on the PMZ in $K_{\rm s}$-band due to the plethora of derived relations in the literature. We selected relations that combined pulsation periods and metallicities to obtain absolute magnitude in $K_{\rm s}$-band.

\begin{figure*}
\includegraphics[width=2\columnwidth]{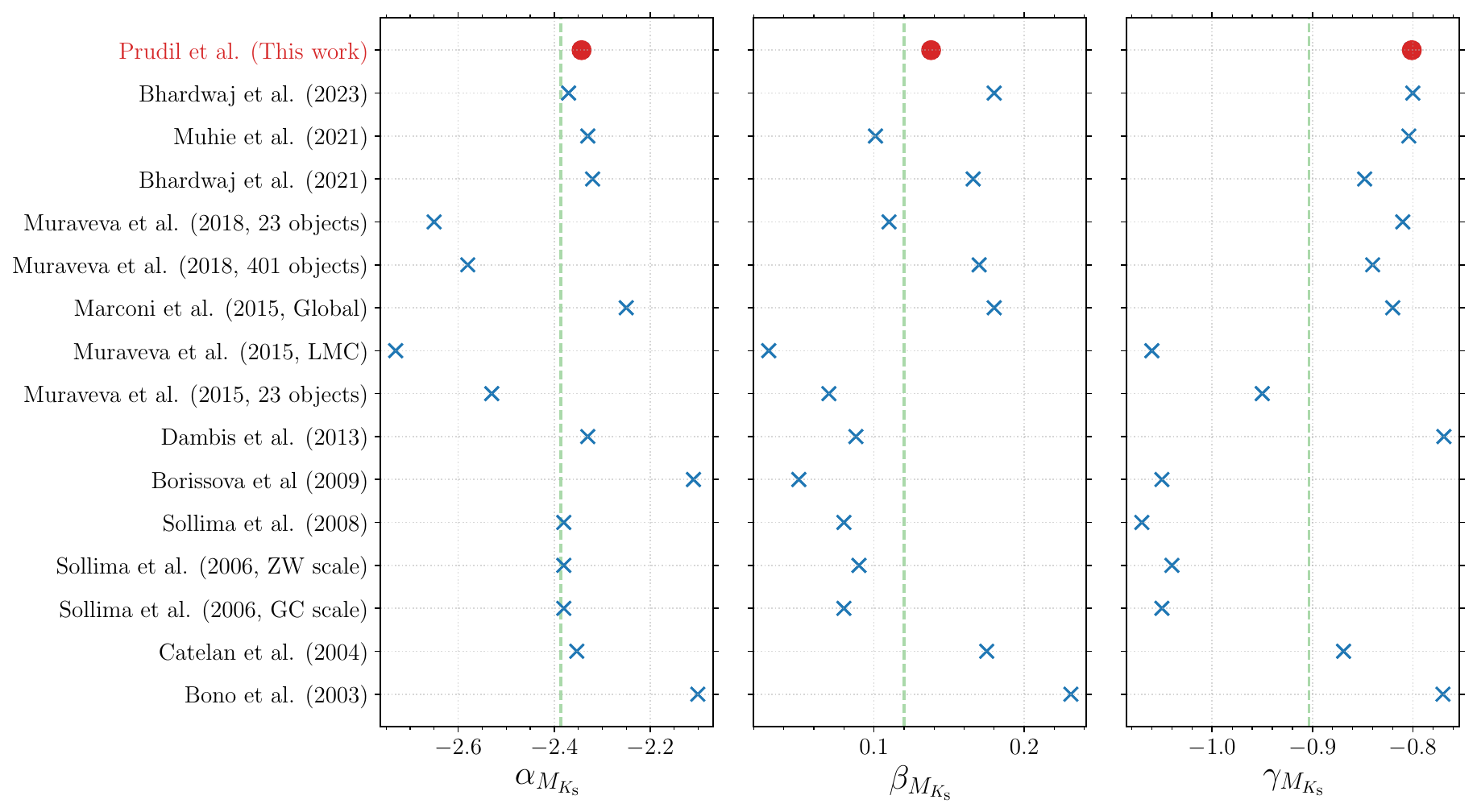}
\caption{Comparison of coefficients for $K_{\rm s}$-passband PMZ relation across different studies \citep[marked with blue crosses, e.g.,][]{Bono2003,Sollima2006,Sollima2008,Borissova2009,Dambis2013,Muraveva2015,Bhardwaj2021,Muhie2021,Bhardwaj2023}. Coefficients derived in this work are marked with red points. The Green dashed line represents the average of coefficients from previous studies.}
\label{fig:PMZ_coef_comparison}
\end{figure*}

In the end, we collected $15$ relations from the literature for the determination of $M_{K_{\rm s}}$, derived from both empirical and theoretical approaches. Figure~\ref{fig:PMZ_coef_comparison} shows a comparison of the literature with the values presented here. Our derived coefficient for the pulsation period, $\alpha_{K_{\rm s}}$, matches quite well with the overall distribution of literature values. A similar conclusion can be made for the metallicity term, $\beta_{K_{\rm s}}$, where despite different metallicity scales used in different studies \citep[e.g.,][]{Zinn1984,Carretta2009} we see a good agreement with the recent studies. Lastly, for the zero-point of the PMZ relations, $\gamma_{K_{\rm s}}$, we see that our derived value agrees exceptionally well with the most recent studies based on \textit{Gaia} parallaxes despite various sources and scales of metallicity for these studies.

\subsection{Comparison with the RR~Lyr star} \label{subsec:ComparsionRRLyr}

We can compare the results presented here with the prototype of the RR~Lyrae class, the RR~Lyr itself, since it did not enter our calibration dataset due to its location at low Galactic latitude ($b = +12.304$\,deg, see criteria in Section~\ref{sec:CalibPMZ}). Using the $Gaia$ DR3 zero-point corrected parallax, we derive a geometrical distance $d = 249.9 \pm 1.9$\,pc \citep[which is in agreement with distances derived by][ with $d = 249.5 \pm 1.5$\,pc]{Bailer2021}. For comparison, the parallax measured by \textit{Hipparcos} \citep{Leeuwen2007cat} yielded a distance of $d = 300 \pm 63$\,pc, and the distance obtained from the parallax measurement by the Hubble Space Telescope was $d = 266 \pm 9$\,pc \citep{Benedict2011}. We used 3D extinction map \citep{Green2019} to obtain a color excess $E(B-V) = 0.01$\,mag\footnote{Using \citet{Schlegel1998} 2D extinction map, we would obtain $E(B-V) = 0.1$\,mag.} toward the RR~Lyr variable. We decided to compare the PMZ predicted distance to the RR~Lyr star only for the $K_{\rm s}$-band PMZ relations since it will mitigate the effects of reddening and $K_{\rm s}$-band PMZ relations are highly numerous in the literature. Therefore, we used the 2MASS mean intensity magnitude, $m_{K_{\rm s}}$, for the RR~Lyr variable and corrected it for pulsation motion using photometric templates from \citet{Braga2019}. As ephemerides for the RR Lyr, we used data from the GEOS database \citep{LeBorgne2007}, with $M_{0} = 2414921.7746$\,day and $P=0.566835616$\,day \citep{LeBorgne2014}. We took into account that different PMZ relations considered different metallicity scales. Thus, for those that used the \citet{Zinn1984} metallicity scale, we used $\text{[Fe/H]}=-1.37$\,dex, while for other relations we used $\text{[Fe/H]}=-1.56$\,dex.

In Table~\ref{tab:RRLyrDistance}, we list a comparison of the distances for the RR~Lyr between different PMZ relations (same as those used in Section~\ref{subsec:ComparsionPrevStudies}). Over half of the PMZ relations overestimate the distance toward the RR~Lyr. The scale of overestimation varies from four parsecs up to $26$\,pc among the different used PMZ laws. Distance differences under five parsecs can be explained by reddening corrections, metallicity variation, and partially by the \textit{Gaia} parallax offset. The difference between the distance from our derived relation and \textit{Gaia} is well within one sigma uncertainty of our distance even if we consider only errors on the observed properties (reddening, metallicity, mean intensity magnitude). Here is important to emphasize the correction of 2MASS mean magnitude for pulsation motion. If we had used the single epoch 2MASS $K_{\rm s}$-band magnitude provided in the 2MASS catalog, we would overestimate the distance toward RR~Lyr in nearly all compared PMZ relations. The distance from the PMZ relation would be shifted by approximately seven parsecs instead.

\begin{table}
\caption{List of distances for the RR~Lyrae star derived using various PMZ relations from the literature. The first column lists references for used PMZ relations. The second column represents the difference between the geometrical distance from $Gaia$ parallax and the distance from the PMZ relation denoted in the third column.}
\footnotesize{
\begin{tabular}{lcc} \hline 
PMZ source & $Gaia - \text{PMZ}$ & distance \\
 & [pc] & [pc]  \\
\hline 
\citet{Bono2003} & $-25.6$ & $275.5$ \\
\citet{Catelan2004} & $-19.2$ & $269.1$ \\
\citet[][GC scale]{Sollima2006} & $-22.5$ & $272.4$ \\
\citet[][ZW scale]{Sollima2006} & $-21.1$ & $271.0$ \\
\citet{Sollima2008} & $-25.0$ & $274.9$ \\
\citet{Borissova2009} & $-25.0$ & $274.9$ \\
\citet{Dambis2013} & \phantom{+0}$9.7$ & $240.2$ \\
\citet[][23 objects]{Muraveva2015} & \phantom{0}$-4.0$ & $253.9$ \\
\citet[][LMC]{Muraveva2015} & \phantom{0}$-3.8$ & $253.7$ \\
\citet[][Global]{Marconi2015} & $-17.3$ & $267.2$ \\
\citet[][400 objects]{Muraveva2018} & \phantom{0}$-4.2$ & $254.1$ \\
\citet[][23 objects]{Muraveva2018} & \phantom{+}$10.5$ & $239.4$ \\
\cite{Bhardwaj2021} & $-15.9$ & $265.8$ \\
\cite{Muhie2021} & \phantom{+0}$3.8$ & $246.1$ \\ 
\cite{Bhardwaj2023} & $-11.3$ & $261.2$ \\ \hline
This work & \phantom{0}$-4.4$ & $254.3$ \\ \hline
\end{tabular}}
\label{tab:RRLyrDistance}
\end{table}

\section{Conclusions} \label{sec:Summary}

In this work, we calibrated empirical relations that describe the dependence of pulsation periods and metallicities on absolute magnitudes for both fundamental and first-overtone RR~Lyrae variables. Our calibration sample was designed to accurately derive PMZ relations for surveys targeting dense stellar structures within the Galaxy. Our study focuses on the OGLE, $Gaia$, and VVV surveys that provide optical and near-infrared photometric observations for dozens of thousands of RR~Lyrae stars in the Galactic disk, bulge, and in the Magellanic Clouds. Despite our calibration sample being smaller than the ones used in the past, we use a homogeneous metallicity scale that can be easily adapted for RR~Lyrae stars in the aforementioned stellar systems.

Our calibration heavily relies on trigonometric parallaxes measured by the $Gaia$ space telescope and the high-resolution metallicities on the CFCS metallicity scale introduced by \citet{Crestani2021}. We considered uncertainties for all involved parameters in our Bayesian approach, including reddening, mean intensity magnitudes, pulsation periods, metallicities, and parallaxes. We also included a parameter for intrinsic scatter in the PMZ in all calibrated passbands to accurately estimate uncertainties in derived absolute magnitudes.

We tested our derived PMZ relations using systems with precise distances and with RR~Lyrae stars with publicly available photometry. For optical $G_{\rm BP}$, near-infrared $I$, $J$, and $K_{\rm s}$ passbands, our relations accurately estimate distance moduli to NGC\,6121, NGC\,5139, LMC, and SMC. In the case of the optical $G_{\rm BP}$ and $G$-bands for the globular cluster NGC\,6121, we detect a significant offset in the derived distance modulus. The offset is mainly caused by the projected reddening and mostly disappears when we use 3D extinction maps toward NGC\,6121.

In addition, we compared our predicted $K_{\rm s}$-band PMZ relation with previously published relations. Our period and metallicity parameters match well with the overall distribution of literature values (even with relations using a different metalicity scale). The zero-point parameter, on the other hand, matches with PMZ relations based on the \textit{Gaia} parallaxes. Moreover, we tested assembled PMZ relation and our derived $K_{\rm s}$-band PMZ to estimate the distance toward the prototype of RR~Lyrae class, the RR~Lyr. We found a good agreement between our derived distance value and distance from \textit{Gaia}. Our derived PMZ relations will be used in the forthcoming papers to examine the structure and kinematics of the Galactic bulge.

\begin{acknowledgements}
Z.P. and A.J.K.H. acknowledge support by the Deutsche Forschungsgemeinschaft (DFG, German Research Foundation) -- Project-ID 138713538 -- SFB 881 (``The Milky Way System", subprojects A03, A05, A11). AMK acknowledges support from grant AST-2009836 from the National Science Foundation. This work has made use of data from the European Space Agency (ESA) mission
{\it Gaia} (\url{https://www.cosmos.esa.int/gaia}), processed by the {\it Gaia}
Data Processing and Analysis Consortium (DPAC,
\url{https://www.cosmos.esa.int/web/gaia/dpac/consortium}). Funding for the DPAC
has been provided by national institutions, in particular the institutions participating in the {\it Gaia} Multilateral Agreement.

This research made use of the following Python packages: \texttt{Astropy} \citep{astropy2013,astropy2018}, \texttt{dustmaps} \citep{Green2018dust}, \texttt{emcee} \citep{Foreman-Mackey2013}, \texttt{IPython} \citep{ipython}, \texttt{Matplotlib} \citep{matplotlib}, \texttt{NumPy} \citep{numpy}, and \texttt{SciPy} \citep{scipy}.
\end{acknowledgements}

\bibliographystyle{aa}
\bibliography{biby}

\begin{thebibliography}{103}
\expandafter\ifx\csname natexlab\endcsname\relax\def\natexlab#1{#1}\fi

\bibitem[{{Astropy Collaboration} {et~al.}(2018){Astropy Collaboration},
  {Price-Whelan}, {Sip{\H{o}}cz}, {G{\"u}nther}, {Lim}, {Crawford}, {Conseil},
  {Shupe}, {Craig}, {Dencheva}, {Ginsburg}, {Vand erPlas}, {Bradley},
  {P{\'e}rez-Su{\'a}rez}, {de Val-Borro}, {Aldcroft}, {Cruz}, {Robitaille},
  {Tollerud}, {Ardelean}, {Babej}, {Bach}, {Bachetti}, {Bakanov}, {Bamford},
  {Barentsen}, {Barmby}, {Baumbach}, {Berry}, {Biscani}, {Boquien}, {Bostroem},
  {Bouma}, {Brammer}, {Bray}, {Breytenbach}, {Buddelmeijer}, {Burke},
  {Calderone}, {Cano Rodr{\'\i}guez}, {Cara}, {Cardoso}, {Cheedella}, {Copin},
  {Corrales}, {Crichton}, {D'Avella}, {Deil}, {Depagne}, {Dietrich}, {Donath},
  {Droettboom}, {Earl}, {Erben}, {Fabbro}, {Ferreira}, {Finethy}, {Fox},
  {Garrison}, {Gibbons}, {Goldstein}, {Gommers}, {Greco}, {Greenfield},
  {Groener}, {Grollier}, {Hagen}, {Hirst}, {Homeier}, {Horton}, {Hosseinzadeh},
  {Hu}, {Hunkeler}, {Ivezi{\'c}}, {Jain}, {Jenness}, {Kanarek}, {Kendrew},
  {Kern}, {Kerzendorf}, {Khvalko}, {King}, {Kirkby}, {Kulkarni}, {Kumar},
  {Lee}, {Lenz}, {Littlefair}, {Ma}, {Macleod}, {Mastropietro}, {McCully},
  {Montagnac}, {Morris}, {Mueller}, {Mumford}, {Muna}, {Murphy}, {Nelson},
  {Nguyen}, {Ninan}, {N{\"o}the}, {Ogaz}, {Oh}, {Parejko}, {Parley}, {Pascual},
  {Patil}, {Patil}, {Plunkett}, {Prochaska}, {Rastogi}, {Reddy Janga},
  {Sabater}, {Sakurikar}, {Seifert}, {Sherbert}, {Sherwood-Taylor}, {Shih},
  {Sick}, {Silbiger}, {Singanamalla}, {Singer}, {Sladen}, {Sooley},
  {Sornarajah}, {Streicher}, {Teuben}, {Thomas}, {Tremblay}, {Turner},
  {Terr{\'o}n}, {van Kerkwijk}, {de la Vega}, {Watkins}, {Weaver}, {Whitmore},
  {Woillez}, {Zabalza}, \& {Astropy Contributors}}]{astropy2018}
{Astropy Collaboration}, {Price-Whelan}, A.~M., {Sip{\H{o}}cz}, B.~M., {et~al.}
  2018, \aj, 156, 123

\bibitem[{{Astropy Collaboration} {et~al.}(2013){Astropy Collaboration},
  {Robitaille}, {Tollerud}, {Greenfield}, {Droettboom}, {Bray}, {Aldcroft},
  {Davis}, {Ginsburg}, {Price-Whelan}, {Kerzendorf}, {Conley}, {Crighton},
  {Barbary}, {Muna}, {Ferguson}, {Grollier}, {Parikh}, {Nair}, {Unther},
  {Deil}, {Woillez}, {Conseil}, {Kramer}, {Turner}, {Singer}, {Fox}, {Weaver},
  {Zabalza}, {Edwards}, {Azalee Bostroem}, {Burke}, {Casey}, {Crawford},
  {Dencheva}, {Ely}, {Jenness}, {Labrie}, {Lim}, {Pierfederici}, {Pontzen},
  {Ptak}, {Refsdal}, {Servillat}, \& {Streicher}}]{astropy2013}
{Astropy Collaboration}, {Robitaille}, T.~P., {Tollerud}, E.~J., {et~al.} 2013,
  \aap, 558, A33

\bibitem[{{Bailer-Jones} {et~al.}(2021){Bailer-Jones}, {Rybizki}, {Fouesneau},
  {Demleitner}, \& {Andrae}}]{Bailer2021}
{Bailer-Jones}, C.~A.~L., {Rybizki}, J., {Fouesneau}, M., {Demleitner}, M., \&
  {Andrae}, R. 2021, \aj, 161, 147

\bibitem[{{Benedict} {et~al.}(2011){Benedict}, {McArthur}, {Feast}, {Barnes},
  {Harrison}, {Bean}, {Menzies}, {Chaboyer}, {Fossati}, {Nesvacil}, {Smith},
  {Kolenberg}, {Laney}, {Kochukhov}, {Nelan}, {Shulyak}, {Taylor}, \&
  {Freedman}}]{Benedict2011}
{Benedict}, G.~F., {McArthur}, B.~E., {Feast}, M.~W., {et~al.} 2011, \aj, 142,
  187

\bibitem[{{Bhardwaj} {et~al.}(2023){Bhardwaj}, {Marconi}, {Rejkuba}, {de
  Grijs}, {Singh}, {Braga}, {Kanbur}, {Ngeow}, {Ripepi}, {Bono}, {De Somma}, \&
  {Dall'Ora}}]{Bhardwaj2023}
{Bhardwaj}, A., {Marconi}, M., {Rejkuba}, M., {et~al.} 2023, \apjl, 944, L51

\bibitem[{{Bhardwaj} {et~al.}(2020){Bhardwaj}, {Rejkuba}, {de Grijs},
  {Herczeg}, {Singh}, {Kanbur}, \& {Ngeow}}]{Bhardwaj2020}
{Bhardwaj}, A., {Rejkuba}, M., {de Grijs}, R., {et~al.} 2020, \aj, 160, 220

\bibitem[{{Bhardwaj} {et~al.}(2021){Bhardwaj}, {Rejkuba}, {de Grijs}, {Yang},
  {Herczeg}, {Marconi}, {Singh}, {Kanbur}, \& {Ngeow}}]{Bhardwaj2021}
{Bhardwaj}, A., {Rejkuba}, M., {de Grijs}, R., {et~al.} 2021, \apj, 909, 200

\bibitem[{{Bono} {et~al.}(1996){Bono}, {Caputo}, {Castellani}, \&
  {Marconi}}]{Bono1996}
{Bono}, G., {Caputo}, F., {Castellani}, V., \& {Marconi}, M. 1996, \apjl, 471,
  L33

\bibitem[{{Bono} {et~al.}(2003){Bono}, {Caputo}, {Castellani}, {Marconi},
  {Storm}, \& {Degl'Innocenti}}]{Bono2003}
{Bono}, G., {Caputo}, F., {Castellani}, V., {et~al.} 2003, \mnras, 344, 1097

\bibitem[{{Borissova} {et~al.}(2009){Borissova}, {Rejkuba}, {Minniti},
  {Catelan}, \& {Ivanov}}]{Borissova2009}
{Borissova}, J., {Rejkuba}, M., {Minniti}, D., {Catelan}, M., \& {Ivanov},
  V.~D. 2009, \aap, 502, 505

\bibitem[{{Braga} {et~al.}(2015){Braga}, {Dall'Ora}, {Bono}, {Stetson},
  {Ferraro}, {Iannicola}, {Marengo}, {Neeley}, {Persson}, {Buonanno},
  {Coppola}, {Freedman}, {Madore}, {Marconi}, {Matsunaga}, {Monson}, {Rich},
  {Scowcroft}, \& {Seibert}}]{Braga2015}
{Braga}, V.~F., {Dall'Ora}, M., {Bono}, G., {et~al.} 2015, \apj, 799, 165

\bibitem[{{Braga} {et~al.}(2016){Braga}, {Stetson}, {Bono}, {Dall'Ora},
  {Ferraro}, {Fiorentino}, {Freyhammer}, {Iannicola}, {Marengo}, {Neeley},
  {Valenti}, {Buonanno}, {Calamida}, {Castellani}, {da Silva},
  {Degl'Innocenti}, {Di Cecco}, {Fabrizio}, {Freedman}, {Giuffrida}, {Lub},
  {Madore}, {Marconi}, {Marinoni}, {Matsunaga}, {Monelli}, {Persson},
  {Piersimoni}, {Pietrinferni}, {Prada-Moroni}, {Pulone}, {Stellingwerf},
  {Tognelli}, \& {Walker}}]{Braga2016}
{Braga}, V.~F., {Stetson}, P.~B., {Bono}, G., {et~al.} 2016, \aj, 152, 170

\bibitem[{{Braga} {et~al.}(2019){Braga}, {Stetson}, {Bono}, {Dall'Ora},
  {Ferraro}, {Fiorentino}, {Iannicola}, {Inno}, {Marengo}, {Neeley}, {Beaton},
  {Buonanno}, {Calamida}, {Contreras Ramos}, {Chaboyer}, {Fabrizio},
  {Freedman}, {Gilligan}, {Johnston}, {Lub}, {Madore}, {Magurno}, {Marconi},
  {Marinoni}, {Marrese}, {Mateo}, {Matsunaga}, {Minniti}, {Monson}, {Monelli},
  {Nonino}, {Persson}, {Pietrinferni}, {Sneden}, {Storm}, {Walker}, {Valenti},
  \& {Zoccali}}]{Braga2019}
{Braga}, V.~F., {Stetson}, P.~B., {Bono}, G., {et~al.} 2019, \aap, 625, A1

\bibitem[{{Braga} {et~al.}(2018){Braga}, {Stetson}, {Bono}, {Dall'Ora},
  {Ferraro}, {Fiorentino}, {Iannicola}, {Marconi}, {Marengo}, {Monson},
  {Neeley}, {Persson}, {Beaton}, {Buonanno}, {Calamida}, {Castellani}, {Di
  Carlo}, {Fabrizio}, {Freedman}, {Inno}, {Madore}, {Magurno}, {Marchetti},
  {Marinoni}, {Marrese}, {Matsunaga}, {Minniti}, {Monelli}, {Nonino},
  {Piersimoni}, {Pietrinferni}, {Prada-Moroni}, {Pulone}, {Stellingwerf},
  {Tognelli}, {Walker}, {Valenti}, \& {Zoccali}}]{Braga2018Ome}
{Braga}, V.~F., {Stetson}, P.~B., {Bono}, G., {et~al.} 2018, \aj, 155, 137

\bibitem[{{Cardelli} {et~al.}(1989){Cardelli}, {Clayton}, \&
  {Mathis}}]{Cardelli1989}
{Cardelli}, J.~A., {Clayton}, G.~C., \& {Mathis}, J.~S. 1989, \apj, 345, 245

\bibitem[{{Carretta} {et~al.}(2009){Carretta}, {Bragaglia}, {Gratton},
  {D'Orazi}, \& {Lucatello}}]{Carretta2009}
{Carretta}, E., {Bragaglia}, A., {Gratton}, R., {D'Orazi}, V., \& {Lucatello},
  S. 2009, \aap, 508, 695

\bibitem[{{Catelan}(2004)}]{Catelan2004Temp}
{Catelan}, M. 2004, \apj, 600, 409

\bibitem[{{Catelan} {et~al.}(2004){Catelan}, {Pritzl}, \&
  {Smith}}]{Catelan2004}
{Catelan}, M., {Pritzl}, B.~J., \& {Smith}, H.~A. 2004, \apjs, 154, 633

\bibitem[{{Chadid} {et~al.}(2017){Chadid}, {Sneden}, \& {Preston}}]{Chadid2017}
{Chadid}, M., {Sneden}, C., \& {Preston}, G.~W. 2017, \apj, 835, 187

\bibitem[{{Clementini} {et~al.}(2022){Clementini}, {Ripepi}, {Garofalo},
  {Molinaro}, {Muraveva}, {Leccia}, {Rimoldini}, {Holl}, {Jevardat de
  Fombelle}, {Sartoretti}, {Marchal}, {Audard}, {Nienartowicz}, {Andrae},
  {Marconi}, {Szabados}, {Evans}, {Lecoeur-Taibi}, {Mowlavi}, {Musella}, \&
  {Eyer}}]{Clementini2022}
{Clementini}, G., {Ripepi}, V., {Garofalo}, A., {et~al.} 2022, arXiv e-prints,
  arXiv:2206.06278

\bibitem[{{Crestani} {et~al.}(2021{\natexlab{a}}){Crestani}, {Braga},
  {Fabrizio}, {Bono}, {Sneden}, {Preston}, {Ferraro}, {Iannicola}, {Nonino},
  {Fiorentino}, {Th{\'e}venin}, {Lemasle}, {Prudil}, {Alves-Brito},
  {Altavilla}, {Chaboyer}, {Dall'Ora}, {D'Orazi}, {Gilligan}, {Grebel},
  {Koch-Hansen}, {Lala}, {Marengo}, {Marinoni}, {Marrese},
  {Mart{\'\i}nez-V{\'a}zquez}, {Matsunaga}, {Monelli}, {Mullen}, {Neeley}, {da
  Silva}, {Stetson}, {Salaris}, {Storm}, {Valenti}, \&
  {Zoccali}}]{Crestani2021Alpha}
{Crestani}, J., {Braga}, V.~F., {Fabrizio}, M., {et~al.} 2021{\natexlab{a}},
  \apj, 914, 10

\bibitem[{{Crestani} {et~al.}(2021{\natexlab{b}}){Crestani}, {Fabrizio},
  {Braga}, {Sneden}, {Preston}, {Ferraro}, {Iannicola}, {Bono}, {Alves-Brito},
  {Nonino}, {D'Orazi}, {Inno}, {Monelli}, {Storm}, {Altavilla}, {Chaboyer},
  {Dall'Ora}, {Fiorentino}, {Gilligan}, {Grebel}, {Lala}, {Lemasle}, {Marengo},
  {Marinoni}, {Marrese}, {Mart{\'\i}nez-V{\'a}zquez}, {Matsunaga}, {Mullen},
  {Neeley}, {Prudil}, {da Silva}, {Stetson}, {Th{\'e}venin}, {Valenti},
  {Walker}, \& {Zoccali}}]{Crestani2021}
{Crestani}, J., {Fabrizio}, M., {Braga}, V.~F., {et~al.} 2021{\natexlab{b}},
  \apj, 908, 20

\bibitem[{{Cusano} {et~al.}(2021){Cusano}, {Moretti}, {Clementini}, {Ripepi},
  {Marconi}, {Cioni}, {Rubele}, {Garofalo}, {de Grijs}, {Groenewegen},
  {Oliveira}, {Subramanian}, {Sun}, \& {van Loon}}]{Cusano2021}
{Cusano}, F., {Moretti}, M.~I., {Clementini}, G., {et~al.} 2021, \mnras, 504, 1

\bibitem[{{Cutri} {et~al.}(2003){Cutri}, {Skrutskie}, {van Dyk}, {Beichman},
  {Carpenter}, {Chester}, {Cambresy}, {Evans}, {Fowler}, {Gizis}, {Howard},
  {Huchra}, {Jarrett}, {Kopan}, {Kirkpatrick}, {Light}, {Marsh}, {McCallon},
  {Schneider}, {Stiening}, {Sykes}, {Weinberg}, {Wheaton}, {Wheelock}, \&
  {Zacarias}}]{Cutri2003}
{Cutri}, R.~M., {Skrutskie}, M.~F., {van Dyk}, S., {et~al.} 2003, VizieR Online
  Data Catalog, II/246

\bibitem[{{Dall'Ora} {et~al.}(2004){Dall'Ora}, {Storm}, {Bono}, {Ripepi},
  {Monelli}, {Testa}, {Andreuzzi}, {Buonanno}, {Caputo}, {Castellani}, {Corsi},
  {Marconi}, {Marconi}, {Pulone}, \& {Stetson}}]{Dallora2004}
{Dall'Ora}, M., {Storm}, J., {Bono}, G., {et~al.} 2004, \apj, 610, 269

\bibitem[{{Dambis}(2009)}]{Dambis2009}
{Dambis}, A.~K. 2009, \mnras, 396, 553

\bibitem[{{Dambis} {et~al.}(2013){Dambis}, {Berdnikov}, {Kniazev}, {Kravtsov},
  {Rastorguev}, {Sefako}, \& {Vozyakova}}]{Dambis2013}
{Dambis}, A.~K., {Berdnikov}, L.~N., {Kniazev}, A.~Y., {et~al.} 2013, \mnras,
  435, 3206

\bibitem[{{D{\'e}k{\'a}ny} {et~al.}(2021){D{\'e}k{\'a}ny}, {Grebel}, \&
  {Pojma{\'n}ski}}]{Dekany2021}
{D{\'e}k{\'a}ny}, I., {Grebel}, E.~K., \& {Pojma{\'n}ski}, G. 2021, \apj, 920,
  33

\bibitem[{{D{\'e}k{\'a}ny} {et~al.}(2013){D{\'e}k{\'a}ny}, {Minniti},
  {Catelan}, {Zoccali}, {Saito}, {Hempel}, \& {Gonzalez}}]{Dekany2013}
{D{\'e}k{\'a}ny}, I., {Minniti}, D., {Catelan}, M., {et~al.} 2013, \apjl, 776,
  L19

\bibitem[{{Du} {et~al.}(2020){Du}, {Mao}, {Athanassoula}, {Shen}, \&
  {Pietrukowicz}}]{Du2020}
{Du}, H., {Mao}, S., {Athanassoula}, E., {Shen}, J., \& {Pietrukowicz}, P.
  2020, \mnras, 498, 5629

\bibitem[{{Fabrizio} {et~al.}(2021){Fabrizio}, {Braga}, {Crestani}, {Bono},
  {Ferraro}, {Fiorentino}, {Iannicola}, {Preston}, {Sneden}, {Th{\'e}venin},
  {Altavilla}, {Chaboyer}, {Dall'Ora}, {da Silva}, {Grebel}, {Gilligan},
  {Lala}, {Lemasle}, {Magurno}, {Marengo}, {Marinoni}, {Marrese},
  {Mart{\'\i}nez-V{\'a}zquez}, {Matsunaga}, {Monelli}, {Mullen}, {Neeley},
  {Nonino}, {Prudil}, {Salaris}, {Stetson}, {Valenti}, \&
  {Zoccali}}]{Fabrizio2021}
{Fabrizio}, M., {Braga}, V.~F., {Crestani}, J., {et~al.} 2021, \apj, 919, 118

\bibitem[{{Fiorentino} {et~al.}(2012){Fiorentino}, {Contreras Ramos},
  {Tolstoy}, {Clementini}, \& {Saha}}]{Fiorentino2012M32}
{Fiorentino}, G., {Contreras Ramos}, R., {Tolstoy}, E., {Clementini}, G., \&
  {Saha}, A. 2012, \aap, 539, A138

\bibitem[{{For} {et~al.}(2011){For}, {Sneden}, \& {Preston}}]{For2011chem}
{For}, B.-Q., {Sneden}, C., \& {Preston}, G.~W. 2011, \apjs, 197, 29

\bibitem[{{Foreman-Mackey} {et~al.}(2013){Foreman-Mackey}, {Hogg}, {Lang}, \&
  {Goodman}}]{Foreman-Mackey2013}
{Foreman-Mackey}, D., {Hogg}, D.~W., {Lang}, D., \& {Goodman}, J. 2013, \pasp,
  125, 306

\bibitem[{{Gaia Collaboration} {et~al.}(2018){Gaia Collaboration}, {Babusiaux},
  {van Leeuwen}, {Barstow}, {Jordi}, {Vallenari}, {Bossini}, {Bressan},
  {Cantat-Gaudin}, {van Leeuwen}, {Brown}, {Prusti}, {de Bruijne},
  {Bailer-Jones}, {Biermann}, {Evans}, {Eyer}, {Jansen}, {Klioner}, {Lammers},
  {Lindegren}, {Luri}, {Mignard}, {Panem}, {Pourbaix}, {Randich}, {Sartoretti},
  {Siddiqui}, {Soubiran}, {Walton}, {Arenou}, {Bastian}, {Cropper}, {Drimmel},
  {Katz}, {Lattanzi}, {Bakker}, {Cacciari}, {Casta{\~n}eda}, {Chaoul}, {Cheek},
  {De Angeli}, {Fabricius}, {Guerra}, {Holl}, {Masana}, {Messineo}, {Mowlavi},
  {Nienartowicz}, {Panuzzo}, {Portell}, {Riello}, {Seabroke}, {Tanga},
  {Th{\'e}venin}, {Gracia-Abril}, {Comoretto}, {Garcia-Reinaldos}, {Teyssier},
  {Altmann}, {Andrae}, {Audard}, {Bellas-Velidis}, {Benson}, {Berthier},
  {Blomme}, {Burgess}, {Busso}, {Carry}, {Cellino}, {Clementini}, {Clotet},
  {Creevey}, {Davidson}, {De Ridder}, {Delchambre}, {Dell'Oro}, {Ducourant},
  {Fern{\'a}ndez-Hern{\'a}ndez}, {Fouesneau}, {Fr{\'e}mat}, {Galluccio},
  {Garc{\'\i}a-Torres}, {Gonz{\'a}lez-N{\'u}{\~n}ez}, {Gonz{\'a}lez-Vidal},
  {Gosset}, {Guy}, {Halbwachs}, {Hambly}, {Harrison}, {Hern{\'a}ndez},
  {Hestroffer}, {Hodgkin}, {Hutton}, {Jasniewicz}, {Jean-Antoine-Piccolo},
  {Jordan}, {Korn}, {Krone-Martins}, {Lanzafame}, {Lebzelter}, {L{\"o}ffler},
  {Manteiga}, {Marrese}, {Mart{\'\i}n-Fleitas}, {Moitinho}, {Mora}, {Muinonen},
  {Osinde}, {Pancino}, {Pauwels}, {Petit}, {Recio-Blanco}, {Richards},
  {Rimoldini}, {Robin}, {Sarro}, {Siopis}, {Smith}, {Sozzetti}, {S{\"u}veges},
  {Torra}, {van Reeven}, {Abbas}, {Abreu Aramburu}, {Accart}, {Aerts},
  {Altavilla}, {{\'A}lvarez}, {Alvarez}, {Alves}, {Anderson}, {Andrei},
  {Anglada Varela}, {Antiche}, {Antoja}, {Arcay}, {Astraatmadja}, {Bach},
  {Baker}, {Balaguer-N{\'u}{\~n}ez}, {Balm}, {Barache}, {Barata}, {Barbato},
  {Barblan}, {Barklem}, {Barrado}, {Barros}, {Bartholom{\'e} Mu{\~n}oz},
  {Bassilana}, {Becciani}, {Bellazzini}, {Berihuete}, {Bertone}, {Bianchi},
  {Bienaym{\'e}}, {Blanco-Cuaresma}, {Boch}, {Boeche}, {Bombrun}, {Borrachero},
  {Bouquillon}, {Bourda}, {Bragaglia}, {Bramante}, {Breddels}, {Brouillet},
  {Br{\"u}semeister}, {Brugaletta}, {Bucciarelli}, {Burlacu}, {Busonero},
  {Butkevich}, {Buzzi}, {Caffau}, {Cancelliere}, {Cannizzaro}, {Carballo},
  {Carlucci}, {Carrasco}, {Casamiquela}, {Castellani}, {Castro-Ginard},
  {Charlot}, {Chemin}, {Chiavassa}, {Cocozza}, {Costigan}, {Cowell}, {Crifo},
  {Crosta}, {Crowley}, {Cuypers}, {Dafonte}, {Damerdji}, {Dapergolas}, {David},
  {David}, {de Laverny}, {De Luise}, {De March}, {de Martino}, {de Souza}, {de
  Torres}, {Debosscher}, {del Pozo}, {Delbo}, {Delgado}, {Delgado}, {Diakite},
  {Diener}, {Distefano}, {Dolding}, {Drazinos}, {Dur{\'a}n}, {Edvardsson},
  {Enke}, {Eriksson}, {Esquej}, {Eynard Bontemps}, {Fabre}, {Fabrizio},
  {Faigler}, {Falc{\~a}o}, {Farr{\`a}s Casas}, {Federici}, {Fedorets},
  {Fernique}, {Figueras}, {Filippi}, {Findeisen}, {Fonti}, {Fraile}, {Fraser},
  {Fr{\'e}zouls}, {Gai}, {Galleti}, {Garabato}, {Garc{\'\i}a-Sedano},
  {Garofalo}, {Garralda}, {Gavel}, {Gavras}, {Gerssen}, {Geyer}, {Giacobbe},
  {Gilmore}, {Girona}, {Giuffrida}, {Glass}, {Gomes}, {Granvik}, {Gueguen},
  {Guerrier}, {Guiraud}, {Guti{\'e}}, {Haigron}, {Hatzidimitriou}, {Hauser},
  {Haywood}, {Heiter}, {Helmi}, {Heu}, {Hilger}, {Hobbs}, {Hofmann}, {Holland},
  {Huckle}, {Hypki}, {Icardi}, {Jan{\ss}en}, {Jevardat de Fombelle}, {Jonker},
  {Juh{\'a}sz}, {Julbe}, {Karampelas}, {Kewley}, {Klar}, {Kochoska}, {Kohley},
  {Kolenberg}, {Kontizas}, {Kontizas}, {Koposov}, {Kordopatis},
  {Kostrzewa-Rutkowska}, {Koubsky}, {Lambert}, {Lanza}, {Lasne}, {Lavigne}, {Le
  Fustec}, {Le Poncin-Lafitte}, {Lebreton}, {Leccia}, {Leclerc},
  {Lecoeur-Taibi}, {Lenhardt}, {Leroux}, {Liao}, {Licata}, {Lindstr{\o}m},
  {Lister}, {Livanou}, {Lobel}, {L{\'o}pez}, {Managau}, {Mann}, {Mantelet},
  {Marchal}, {Marchant}, {Marconi}, {Marinoni}, {Marschalk{\'o}}, {Marshall},
  {Martino}, {Marton}, {Mary}, {Massari}, {Matijevi{\v{c}}}, {Mazeh},
  {McMillan}, {Messina}, {Michalik}, {Millar}, {Molina}, {Molinaro},
  {Moln{\'a}r}, {Montegriffo}, {Mor}, {Morbidelli}, {Morel}, {Morris},
  {Mulone}, {Muraveva}, {Musella}, {Nelemans}, {Nicastro}, {Noval},
  {O'Mullane}, {Ord{\'e}novic}, {Ord{\'o}{\~n}ez-Blanco}, {Osborne}, {Pagani},
  {Pagano}, {Pailler}, {Palacin}, {Palaversa}, {Panahi}, {Pawlak},
  {Piersimoni}, {Pineau}, {Plachy}, {Plum}, {Poggio}, {Poujoulet},
  {Pr{\v{s}}a}, {Pulone}, {Racero}, {Ragaini}, {Rambaux}, {Ramos-Lerate},
  {Regibo}, {Reyl{\'e}}, {Riclet}, {Ripepi}, {Riva}, {Rivard}, {Rixon},
  {Roegiers}, {Roelens}, {Romero-G{\'o}mez}, {Rowell}, {Royer}, {Ruiz-Dern},
  {Sadowski}, {Sagrist{\`a} Sell{\'e}s}, {Sahlmann}, {Salgado}, {Salguero},
  {Sanna}, {Santana-Ros}, {Sarasso}, {Savietto}, {Schultheis}, {Sciacca},
  {Segol}, {Segovia}, {S{\'e}gransan}, {Shih}, {Siltala}, {Silva}, {Smart},
  {Smith}, {Solano}, {Solitro}, {Sordo}, {Soria Nieto}, {Souchay}, {Spagna},
  {Spoto}, {Stampa}, {Steele}, {Steidelm{\"u}ller}, {Stephenson}, {Stoev},
  {Suess}, {Surdej}, {Szabados}, {Szegedi-Elek}, {Tapiador}, {Taris}, {Tauran},
  {Taylor}, {Teixeira}, {Terrett}, {Teyssandier}, {Thuillot}, {Titarenko},
  {Torra Clotet}, {Turon}, {Ulla}, {Utrilla}, {Uzzi}, {Vaillant}, {Valentini},
  {Valette}, {van Elteren}, {Van Hemelryck}, {Vaschetto}, {Vecchiato},
  {Veljanoski}, {Viala}, {Vicente}, {Vogt}, {von Essen}, {Voss}, {Votruba},
  {Voutsinas}, {Walmsley}, {Weiler}, {Wertz}, {Wevers}, {Wyrzykowski},
  {Yoldas}, {{\v{Z}}erjal}, {Ziaeepour}, {Zorec}, {Zschocke}, {Zucker},
  {Zurbach}, \& {Zwitter}}]{Hertzsprung2018ExtRed}
{Gaia Collaboration}, {Babusiaux}, C., {van Leeuwen}, F., {et~al.} 2018, \aap,
  616, A10

\bibitem[{{Gaia Collaboration} {et~al.}(2016){Gaia Collaboration}, {Prusti},
  {de Bruijne}, {Brown}, {Vallenari}, {Babusiaux}, {Bailer-Jones}, {Bastian},
  {Biermann}, {Evans}, {Eyer}, {Jansen}, {Jordi}, {Klioner}, {Lammers},
  {Lindegren}, {Luri}, {Mignard}, {Milligan}, {Panem}, {Poinsignon},
  {Pourbaix}, {Randich}, {Sarri}, {Sartoretti}, {Siddiqui}, {Soubiran},
  {Valette}, {van Leeuwen}, {Walton}, {Aerts}, {Arenou}, {Cropper}, {Drimmel},
  {H{\o}g}, {Katz}, {Lattanzi}, {O'Mullane}, {Grebel}, {Holland}, {Huc},
  {Passot}, {Bramante}, {Cacciari}, {Casta{\~n}eda}, {Chaoul}, {Cheek}, {De
  Angeli}, {Fabricius}, {Guerra}, {Hern{\'a}ndez}, {Jean-Antoine-Piccolo},
  {Masana}, {Messineo}, {Mowlavi}, {Nienartowicz}, {Ord{\'o}{\~n}ez-Blanco},
  {Panuzzo}, {Portell}, {Richards}, {Riello}, {Seabroke}, {Tanga},
  {Th{\'e}venin}, {Torra}, {Els}, {Gracia-Abril}, {Comoretto},
  {Garcia-Reinaldos}, {Lock}, {Mercier}, {Altmann}, {Andrae}, {Astraatmadja},
  {Bellas-Velidis}, {Benson}, {Berthier}, {Blomme}, {Busso}, {Carry},
  {Cellino}, {Clementini}, {Cowell}, {Creevey}, {Cuypers}, {Davidson}, {De
  Ridder}, {de Torres}, {Delchambre}, {Dell'Oro}, {Ducourant}, {Fr{\'e}mat},
  {Garc{\'\i}a-Torres}, {Gosset}, {Halbwachs}, {Hambly}, {Harrison}, {Hauser},
  {Hestroffer}, {Hodgkin}, {Huckle}, {Hutton}, {Jasniewicz}, {Jordan},
  {Kontizas}, {Korn}, {Lanzafame}, {Manteiga}, {Moitinho}, {Muinonen},
  {Osinde}, {Pancino}, {Pauwels}, {Petit}, {Recio-Blanco}, {Robin}, {Sarro},
  {Siopis}, {Smith}, {Smith}, {Sozzetti}, {Thuillot}, {van Reeven}, {Viala},
  {Abbas}, {Abreu Aramburu}, {Accart}, {Aguado}, {Allan}, {Allasia},
  {Altavilla}, {{\'A}lvarez}, {Alves}, {Anderson}, {Andrei}, {Anglada Varela},
  {Antiche}, {Antoja}, {Ant{\'o}n}, {Arcay}, {Atzei}, {Ayache}, {Bach},
  {Baker}, {Balaguer-N{\'u}{\~n}ez}, {Barache}, {Barata}, {Barbier}, {Barblan},
  {Baroni}, {Barrado y Navascu{\'e}s}, {Barros}, {Barstow}, {Becciani},
  {Bellazzini}, {Bellei}, {Bello Garc{\'\i}a}, {Belokurov}, {Bendjoya},
  {Berihuete}, {Bianchi}, {Bienaym{\'e}}, {Billebaud}, {Blagorodnova},
  {Blanco-Cuaresma}, {Boch}, {Bombrun}, {Borrachero}, {Bouquillon}, {Bourda},
  {Bouy}, {Bragaglia}, {Breddels}, {Brouillet}, {Br{\"u}semeister},
  {Bucciarelli}, {Budnik}, {Burgess}, {Burgon}, {Burlacu}, {Busonero}, {Buzzi},
  {Caffau}, {Cambras}, {Campbell}, {Cancelliere}, {Cantat-Gaudin}, {Carlucci},
  {Carrasco}, {Castellani}, {Charlot}, {Charnas}, {Charvet}, {Chassat},
  {Chiavassa}, {Clotet}, {Cocozza}, {Collins}, {Collins}, {Costigan}, {Crifo},
  {Cross}, {Crosta}, {Crowley}, {Dafonte}, {Damerdji}, {Dapergolas}, {David},
  {David}, {De Cat}, {de Felice}, {de Laverny}, {De Luise}, {De March}, {de
  Martino}, {de Souza}, {Debosscher}, {del Pozo}, {Delbo}, {Delgado},
  {Delgado}, {di Marco}, {Di Matteo}, {Diakite}, {Distefano}, {Dolding}, {Dos
  Anjos}, {Drazinos}, {Dur{\'a}n}, {Dzigan}, {Ecale}, {Edvardsson}, {Enke},
  {Erdmann}, {Escolar}, {Espina}, {Evans}, {Eynard Bontemps}, {Fabre},
  {Fabrizio}, {Faigler}, {Falc{\~a}o}, {Farr{\`a}s Casas}, {Faye}, {Federici},
  {Fedorets}, {Fern{\'a}ndez-Hern{\'a}ndez}, {Fernique}, {Fienga}, {Figueras},
  {Filippi}, {Findeisen}, {Fonti}, {Fouesneau}, {Fraile}, {Fraser}, {Fuchs},
  {Furnell}, {Gai}, {Galleti}, {Galluccio}, {Garabato}, {Garc{\'\i}a-Sedano},
  {Gar{\'e}}, {Garofalo}, {Garralda}, {Gavras}, {Gerssen}, {Geyer}, {Gilmore},
  {Girona}, {Giuffrida}, {Gomes}, {Gonz{\'a}lez-Marcos},
  {Gonz{\'a}lez-N{\'u}{\~n}ez}, {Gonz{\'a}lez-Vidal}, {Granvik}, {Guerrier},
  {Guillout}, {Guiraud}, {G{\'u}rpide}, {Guti{\'e}rrez-S{\'a}nchez}, {Guy},
  {Haigron}, {Hatzidimitriou}, {Haywood}, {Heiter}, {Helmi}, {Hobbs},
  {Hofmann}, {Holl}, {Holland }, {Hunt}, {Hypki}, {Icardi}, {Irwin}, {Jevardat
  de Fombelle}, {Jofr{\'e}}, {Jonker}, {Jorissen}, {Julbe}, {Karampelas},
  {Kochoska}, {Kohley}, {Kolenberg}, {Kontizas}, {Koposov}, {Kordopatis},
  {Koubsky}, {Kowalczyk}, {Krone-Martins}, {Kudryashova}, {Kull}, {Bachchan},
  {Lacoste-Seris}, {Lanza}, {Lavigne}, {Le Poncin-Lafitte}, {Lebreton},
  {Lebzelter}, {Leccia}, {Leclerc}, {Lecoeur-Taibi}, {Lemaitre}, {Lenhardt},
  {Leroux}, {Liao}, {Licata}, {Lindstr{\o}m}, {Lister}, {Livanou}, {Lobel},
  {L{\"o}ffler}, {L{\'o}pez}, {Lopez-Lozano}, {Lorenz}, {Loureiro},
  {MacDonald}, {Magalh{\~a}es Fernandes}, {Managau}, {Mann}, {Mantelet},
  {Marchal}, {Marchant}, {Marconi}, {Marie}, {Marinoni}, {Marrese},
  {Marschalk{\'o}}, {Marshall}, {Mart{\'\i}n-Fleitas}, {Martino}, {Mary},
  {Matijevi{\v{c}}}, {Mazeh}, {McMillan}, {Messina}, {Mestre}, {Michalik},
  {Millar}, {Miranda}, {Molina}, {Molinaro}, {Molinaro}, {Moln{\'a}r},
  {Moniez}, {Montegriffo}, {Monteiro}, {Mor}, {Mora}, {Morbidelli}, {Morel},
  {Morgenthaler}, {Morley}, {Morris}, {Mulone}, {Muraveva}, {Musella},
  {Narbonne}, {Nelemans}, {Nicastro}, {Noval}, {Ord{\'e}novic},
  {Ordieres-Mer{\'e}}, {Osborne}, {Pagani}, {Pagano}, {Pailler}, {Palacin},
  {Palaversa}, {Parsons}, {Paulsen}, {Pecoraro}, {Pedrosa}, {Pentik{\"a}inen},
  {Pereira}, {Pichon}, {Piersimoni}, {Pineau}, {Plachy}, {Plum}, {Poujoulet},
  {Pr{\v{s}}a}, {Pulone}, {Ragaini}, {Rago}, {Rambaux}, {Ramos-Lerate},
  {Ranalli}, {Rauw}, {Read}, {Regibo}, {Renk}, {Reyl{\'e}}, {Ribeiro},
  {Rimoldini}, {Ripepi}, {Riva}, {Rixon}, {Roelens}, {Romero-G{\'o}mez},
  {Rowell}, {Royer}, {Rudolph}, {Ruiz-Dern}, {Sadowski}, {Sagrist{\`a}
  Sell{\'e}s}, {Sahlmann}, {Salgado}, {Salguero}, {Sarasso}, {Savietto},
  {Schnorhk}, {Schultheis}, {Sciacca}, {Segol}, {Segovia}, {Segransan},
  {Serpell}, {Shih}, {Smareglia}, {Smart}, {Smith}, {Solano}, {Solitro},
  {Sordo}, {Soria Nieto}, {Souchay}, {Spagna}, {Spoto}, {Stampa}, {Steele},
  {Steidelm{\"u}ller}, {Stephenson}, {Stoev}, {Suess}, {S{\"u}veges}, {Surdej},
  {Szabados}, {Szegedi-Elek}, {Tapiador}, {Taris}, {Tauran}, {Taylor},
  {Teixeira}, {Terrett}, {Tingley}, {Trager}, {Turon}, {Ulla}, {Utrilla},
  {Valentini}, {van Elteren}, {Van Hemelryck}, {van Leeuwen}, {Varadi},
  {Vecchiato}, {Veljanoski}, {Via}, {Vicente}, {Vogt}, {Voss}, {Votruba},
  {Voutsinas}, {Walmsley}, {Weiler}, {Weingrill}, {Werner}, {Wevers},
  {Whitehead}, {Wyrzykowski}, {Yoldas}, {{\v{Z}}erjal}, {Zucker}, {Zurbach},
  {Zwitter}, {Alecu}, {Allen}, {Allende Prieto}, {Amorim},
  {Anglada-Escud{\'e}}, {Arsenijevic}, {Azaz}, {Balm}, {Beck}, {Bernstein},
  {Bigot}, {Bijaoui}, {Blasco}, {Bonfigli}, {Bono}, {Boudreault}, {Bressan},
  {Brown}, {Brunet}, {Bunclark}, {Buonanno}, {Butkevich}, {Carret}, {Carrion},
  {Chemin}, {Ch{\'e}reau}, {Corcione}, {Darmigny}, {de Boer}, {de Teodoro}, {de
  Zeeuw}, {Delle Luche}, {Domingues}, {Dubath}, {Fodor}, {Fr{\'e}zouls},
  {Fries}, {Fustes}, {Fyfe}, {Gallardo}, {Gallegos}, {Gardiol}, {Gebran},
  {Gomboc}, {G{\'o}mez}, {Grux}, {Gueguen}, {Heyrovsky}, {Hoar}, {Iannicola},
  {Isasi Parache}, {Janotto}, {Joliet}, {Jonckheere}, {Keil}, {Kim},
  {Klagyivik}, {Klar}, {Knude}, {Kochukhov}, {Kolka}, {Kos}, {Kutka}, {Lainey},
  {LeBouquin}, {Liu}, {Loreggia}, {Makarov}, {Marseille}, {Martayan},
  {Martinez-Rubi}, {Massart}, {Meynadier}, {Mignot}, {Munari}, {Nguyen},
  {Nordlander}, {Ocvirk}, {O'Flaherty}, {Olias Sanz}, {Ortiz}, {Osorio},
  {Oszkiewicz}, {Ouzounis}, {Palmer}, {Park}, {Pasquato}, {Peltzer}, {Peralta},
  {P{\'e}turaud}, {Pieniluoma}, {Pigozzi}, {Poels}, {Prat}, {Prod'homme},
  {Raison}, {Rebordao}, {Risquez}, {Rocca-Volmerange}, {Rosen}, {Ruiz-Fuertes},
  {Russo}, {Sembay}, {Serraller Vizcaino}, {Short}, {Siebert}, {Silva},
  {Sinachopoulos}, {Slezak}, {Soffel}, {Sosnowska}, {Strai{\v{z}}ys}, {ter
  Linden}, {Terrell}, {Theil}, {Tiede}, {Troisi}, {Tsalmantza}, {Tur},
  {Vaccari}, {Vachier}, {Valles}, {Van Hamme}, {Veltz}, {Virtanen}, {Wallut},
  {Wichmann}, {Wilkinson}, {Ziaeepour}, \& {Zschocke}}]{Gaia2016}
{Gaia Collaboration}, {Prusti}, T., {de Bruijne}, J.~H.~J., {et~al.} 2016,
  \aap, 595, A1

\bibitem[{{Gaia Collaboration} {et~al.}(2022){Gaia Collaboration}, {Vallenari},
  {Brown}, {Prusti}, {de Bruijne}, {Arenou}, {Babusiaux}, {Biermann},
  {Creevey}, {Ducourant}, \& et~al.}]{GaiaDR32022}
{Gaia Collaboration}, {Vallenari}, A., {Brown}, A.~G.~A., {et~al.} 2022, arXiv
  e-prints, arXiv:2208.00211

\bibitem[{{Garofalo} {et~al.}(2022){Garofalo}, {Delgado}, {Sarro},
  {Clementini}, {Muraveva}, {Marconi}, \& {Ripepi}}]{Garofalo2022}
{Garofalo}, A., {Delgado}, H.~E., {Sarro}, L.~M., {et~al.} 2022, \mnras, 513,
  788

\bibitem[{{Graczyk} {et~al.}(2020){Graczyk}, {Pietrzy{\'n}ski}, {Thompson},
  {Gieren}, {Zgirski}, {Villanova}, {G{\'o}rski}, {Wielg{\'o}rski},
  {Karczmarek}, {Narloch}, {Pilecki}, {Taormina}, {Smolec}, {Suchomska},
  {Gallenne}, {Nardetto}, {Storm}, {Kudritzki}, {Ka{\l}uszy{\'n}ski}, \&
  {Pych}}]{Graczyk2020SMC}
{Graczyk}, D., {Pietrzy{\'n}ski}, G., {Thompson}, I.~B., {et~al.} 2020, \apj,
  904, 13

\bibitem[{{Green}(2018)}]{Green2018dust}
{Green}, G. 2018, The Journal of Open Source Software, 3, 695

\bibitem[{{Green} {et~al.}(2019){Green}, {Schlafly}, {Zucker}, {Speagle}, \&
  {Finkbeiner}}]{Green2019}
{Green}, G.~M., {Schlafly}, E., {Zucker}, C., {Speagle}, J.~S., \&
  {Finkbeiner}, D. 2019, \apj, 887, 93

\bibitem[{Harris {et~al.}(2020)Harris, Millman, van~der Walt, Gommers,
  Virtanen, Cournapeau, Wieser, Taylor, Berg, Smith, Kern, Picus, Hoyer, van
  Kerkwijk, Brett, Haldane, Fernández~del Río, Wiebe, Peterson,
  Gérard-Marchant, Sheppard, Reddy, Weckesser, Abbasi, Gohlke, \&
  Oliphant}]{numpy}
Harris, C.~R., Millman, K.~J., van~der Walt, S.~J., {et~al.} 2020, Nature, 585,
  357–362

\bibitem[{Hunter(2007)}]{matplotlib}
Hunter, J.~D. 2007, Computing in Science \& Engineering, 9, 90

\bibitem[{{Iben} \& {Huchra}(1971)}]{Iben1971}
{Iben}, I., J. \& {Huchra}, J. 1971, \aap, 14, 293

\bibitem[{{Jacyszyn-Dobrzeniecka} {et~al.}(2020){Jacyszyn-Dobrzeniecka},
  {Mr{\'o}z}, {Kruszy{\'n}ska}, {Soszy{\'n}ski}, {Skowron}, {Udalski},
  {Szyma{\'n}ski}, {Iwanek}, {Skowron}, {Pietrukowicz}, {Poleski},
  {Koz{\l}owski}, {Ulaczyk}, {Rybicki}, \& {Wrona}}]{Jacyszyn-Dobrzeniecka2020}
{Jacyszyn-Dobrzeniecka}, A.~M., {Mr{\'o}z}, P., {Kruszy{\'n}ska}, K., {et~al.}
  2020, \apj, 889, 26

\bibitem[{{Jacyszyn-Dobrzeniecka} {et~al.}(2017){Jacyszyn-Dobrzeniecka},
  {Skowron}, {Mr{\'o}z}, {Soszy{\'n}ski}, {Udalski}, {Pietrukowicz}, {Skowron},
  {Poleski}, {Koz{\l}owski}, {Wyrzykowski}, {Pawlak}, {Szyma{\'n}ski}, \&
  {Ulaczyk}}]{Jacyszyn-Dobrzeniecka2017}
{Jacyszyn-Dobrzeniecka}, A.~M., {Skowron}, D.~M., {Mr{\'o}z}, P., {et~al.}
  2017, \actaa, 67, 1

\bibitem[{{Jayasinghe} {et~al.}(2018){Jayasinghe}, {Kochanek}, {Stanek},
  {Shappee}, {Holoien}, {Thompson}, {Prieto}, {Dong}, {Pawlak}, {Shields},
  {Pojmanski}, {Otero}, {Britt}, \& {Will}}]{Jayasinghe2018}
{Jayasinghe}, T., {Kochanek}, C.~S., {Stanek}, K.~Z., {et~al.} 2018, \mnras,
  477, 3145

\bibitem[{Jaynes(1968)}]{Jaynes1968}
Jaynes, E.~T. 1968, IEEE Transactions on Systems Science and Cybernetics, 4,
  227

\bibitem[{{Kunder} {et~al.}(2020){Kunder}, {P{\'e}rez-Villegas}, {Rich},
  {Ogata}, {Murari}, {Boren}, {Johnson}, {Nataf}, {Walker}, {Bono}, {Koch},
  {Propris}, {Storm}, \& {Wojno}}]{Kunder2020}
{Kunder}, A., {P{\'e}rez-Villegas}, A., {Rich}, R.~M., {et~al.} 2020, \aj, 159,
  270

\bibitem[{{Kunder} {et~al.}(2016){Kunder}, {Rich}, {Koch}, {Storm}, {Nataf},
  {De Propris}, {Walker}, {Bono}, {Johnson}, {Shen}, \& {Li}}]{Kunder2016}
{Kunder}, A., {Rich}, R.~M., {Koch}, A., {et~al.} 2016, \apjl, 821, L25

\bibitem[{{Layden} {et~al.}(2019){Layden}, {Tiede}, {Chaboyer}, {Bunner}, \&
  {Smitka}}]{Layden2019}
{Layden}, A.~C., {Tiede}, G.~P., {Chaboyer}, B., {Bunner}, C., \& {Smitka},
  M.~T. 2019, \aj, 158, 105

\bibitem[{{Le Borgne} {et~al.}(2007){Le Borgne}, {Paschke}, {Vandenbroere},
  {Poretti}, {Klotz}, {Bo{\"e}r}, {Damerdji}, {Martignoni}, \&
  {Acerbi}}]{LeBorgne2007}
{Le Borgne}, J.~F., {Paschke}, A., {Vandenbroere}, J., {et~al.} 2007, \aap,
  476, 307

\bibitem[{{Lindegren} {et~al.}(2021){Lindegren}, {Bastian}, {Biermann},
  {Bombrun}, {de Torres}, {Gerlach}, {Geyer}, {Hern{\'a}ndez}, {Hilger},
  {Hobbs}, {Klioner}, {Lammers}, {McMillan}, {Ramos-Lerate},
  {Steidelm{\"u}ller}, {Stephenson}, \& {van Leeuwen}}]{Lindegren2021}
{Lindegren}, L., {Bastian}, U., {Biermann}, M., {et~al.} 2021, \aap, 649, A4

\bibitem[{{Longmore} {et~al.}(1986){Longmore}, {Fernley}, \&
  {Jameson}}]{Longmore1986}
{Longmore}, A.~J., {Fernley}, J.~A., \& {Jameson}, R.~F. 1986, \mnras, 220, 279

\bibitem[{{Marconi} {et~al.}(2018){Marconi}, {Bono}, {Pietrinferni}, {Braga},
  {Castellani}, \& {Stellingwerf}}]{Marconi2018PLZ}
{Marconi}, M., {Bono}, G., {Pietrinferni}, A., {et~al.} 2018, \apjl, 864, L13

\bibitem[{{Marconi} {et~al.}(2015){Marconi}, {Coppola}, {Bono}, {Braga},
  {Pietrinferni}, {Buonanno}, {Castellani}, {Musella}, {Ripepi}, \&
  {Stellingwerf}}]{Marconi2015}
{Marconi}, M., {Coppola}, G., {Bono}, G., {et~al.} 2015, \apj, 808, 50

\bibitem[{{Mart{\'\i}nez-V{\'a}zquez}
  {et~al.}(2016){Mart{\'\i}nez-V{\'a}zquez}, {Stetson}, {Monelli}, {Bernard},
  {Fiorentino}, {Gallart}, {Bono}, {Cassisi}, {Dall'Ora}, {Ferraro},
  {Iannicola}, \& {Walker}}]{MartinezVazquez2016}
{Mart{\'\i}nez-V{\'a}zquez}, C.~E., {Stetson}, P.~B., {Monelli}, M., {et~al.}
  2016, \mnras, 462, 4349

\bibitem[{{Mart{\'\i}nez-V{\'a}zquez}
  {et~al.}(2019){Mart{\'\i}nez-V{\'a}zquez}, {Vivas}, {Gurevich}, {Walker},
  {McCarthy}, {Pace}, {Stringer}, {Santiago}, {Hounsell}, {Macri}, {Li},
  {Bechtol}, {Riley}, {Kim}, {Simon}, {Drlica-Wagner}, {Nadler}, {Marshall},
  {Annis}, {Avila}, {Bertin}, {Brooks}, {Buckley-Geer}, {Burke}, {Carnero
  Rosell}, {Carrasco Kind}, {da Costa}, {De Vicente}, {Desai}, {Diehl}, {Doel},
  {Everett}, {Frieman}, {Garc{\'\i}a-Bellido}, {Gaztanaga}, {Gruen}, {Gruendl},
  {Gschwend}, {Gutierrez}, {Hollowood}, {Honscheid}, {James}, {Kuehn},
  {Kuropatkin}, {Maia}, {Menanteau}, {Miller}, {Miquel}, {Paz-Chinch{\'o}n},
  {Plazas}, {Sanchez}, {Scarpine}, {Serrano}, {Sevilla-Noarbe}, {Smith},
  {Soares-Santos}, {Sobreira}, {Swanson}, {Tarle}, {Vikram}, \& {DES
  Collaboration}}]{MartinezVazquez2019}
{Mart{\'\i}nez-V{\'a}zquez}, C.~E., {Vivas}, A.~K., {Gurevich}, M., {et~al.}
  2019, \mnras, 490, 2183

\bibitem[{{Minniti} {et~al.}(2010){Minniti}, {Lucas}, {Emerson}, {Saito},
  {Hempel}, {Pietrukowicz}, {Ahumada}, {Alonso}, {Alonso-Garcia}, {Arias},
  {Bandyopadhyay}, {Barb{\'a}}, {Barbuy}, {Bedin}, {Bica}, {Borissova},
  {Bronfman}, {Carraro}, {Catelan}, {Clari{\'a}}, {Cross}, {de Grijs},
  {D{\'e}k{\'a}ny}, {Drew}, {Fari{\~n}a}, {Feinstein}, {Fern{\'a}ndez
  Laj{\'u}s}, {Gamen}, {Geisler}, {Gieren}, {Goldman}, {Gonzalez}, {Gunthardt},
  {Gurovich}, {Hambly}, {Irwin}, {Ivanov}, {Jord{\'a}n}, {Kerins}, {Kinemuchi},
  {Kurtev}, {L{\'o}pez-Corredoira}, {Maccarone}, {Masetti}, {Merlo},
  {Messineo}, {Mirabel}, {Monaco}, {Morelli}, {Padilla}, {Palma}, {Parisi},
  {Pignata}, {Rejkuba}, {Roman-Lopes}, {Sale}, {Schreiber}, {Schr{\"o}der},
  {Smith}, {}, {Soto}, {Tamura}, {Tappert}, {Thompson}, {Toledo}, {Zoccali}, \&
  {Pietrzynski}}]{Minniti2010}
{Minniti}, D., {Lucas}, P.~W., {Emerson}, J.~P., {et~al.} 2010, \na, 15, 433

\bibitem[{{Moln{\'a}r} {et~al.}(2022){Moln{\'a}r}, {B{\'o}di}, {P{\'a}l},
  {Bhardwaj}, {Hambsch}, {Benk{\H{o}}}, {Derekas}, {Ebadi}, {Joyce},
  {Hasanzadeh}, {Kolenberg}, {Lund}, {Nemec}, {Netzel}, {Ngeow}, {Pepper},
  {Plachy}, {Prudil}, {Siverd}, {Skarka}, {Smolec}, {S{\'o}dor}, {Sylla},
  {Szab{\'o}}, {Szab{\'o}}, {Kjeldsen}, {Christensen-Dalsgaard}, \&
  {Ricker}}]{Molnar2022RRlyr}
{Moln{\'a}r}, L., {B{\'o}di}, A., {P{\'a}l}, A., {et~al.} 2022, \apjs, 258, 8

\bibitem[{{Molnar} {et~al.}(2022){Molnar}, {Sanders}, {Smith}, {Belokurov},
  {Lucas}, \& {Minniti}}]{Molnar2022}
{Molnar}, T.~A., {Sanders}, J.~L., {Smith}, L.~C., {et~al.} 2022, \mnras, 509,
  2566

\bibitem[{{Muhie} {et~al.}(2021){Muhie}, {Dambis}, {Berdnikov}, {Kniazev}, \&
  {Grebel}}]{Muhie2021}
{Muhie}, T.~D., {Dambis}, A.~K., {Berdnikov}, L.~N., {Kniazev}, A.~Y., \&
  {Grebel}, E.~K. 2021, \mnras, 502, 4074

\bibitem[{{Muraveva} {et~al.}(2018{\natexlab{a}}){Muraveva}, {Delgado},
  {Clementini}, {Sarro}, \& {Garofalo}}]{Muraveva2018}
{Muraveva}, T., {Delgado}, H.~E., {Clementini}, G., {Sarro}, L.~M., \&
  {Garofalo}, A. 2018{\natexlab{a}}, \mnras, 481, 1195

\bibitem[{{Muraveva} {et~al.}(2015){Muraveva}, {Palmer}, {Clementini}, {Luri},
  {Cioni}, {Moretti}, {Marconi}, {Ripepi}, \& {Rubele}}]{Muraveva2015}
{Muraveva}, T., {Palmer}, M., {Clementini}, G., {et~al.} 2015, \apj, 807, 127

\bibitem[{{Muraveva} {et~al.}(2018{\natexlab{b}}){Muraveva}, {Subramanian},
  {Clementini}, {Cioni}, {Palmer}, {van Loon}, {Moretti}, {de Grijs},
  {Molinaro}, {Ripepi}, {Marconi}, {Emerson}, \& {Ivanov}}]{Muraveva2018SMC}
{Muraveva}, T., {Subramanian}, S., {Clementini}, G., {et~al.}
  2018{\natexlab{b}}, \mnras, 473, 3131

\bibitem[{{Neeley} {et~al.}(2017){Neeley}, {Marengo}, {Bono}, {Braga},
  {Dall'Ora}, {Magurno}, {Marconi}, {Trueba}, {Tognelli}, {Prada Moroni},
  {Beaton}, {Freedman}, {Madore}, {Monson}, {Scowcroft}, {Seibert}, \&
  {Stetson}}]{Neeley2017}
{Neeley}, J.~R., {Marengo}, M., {Bono}, G., {et~al.} 2017, \apj, 841, 84

\bibitem[{{Neeley} {et~al.}(2015){Neeley}, {Marengo}, {Bono}, {Braga},
  {Dall'Ora}, {Stetson}, {Buonanno}, {Ferraro}, {Freedman}, {Iannicola},
  {Madore}, {Matsunaga}, {Monson}, {Persson}, {Scowcroft}, \&
  {Seibert}}]{Neeley2015}
{Neeley}, J.~R., {Marengo}, M., {Bono}, G., {et~al.} 2015, \apj, 808, 11

\bibitem[{{Neeley} {et~al.}(2019){Neeley}, {Marengo}, {Freedman}, {Madore},
  {Beaton}, {Hatt}, {Hoyt}, {Monson}, {Rich}, {Sarajedini}, {Seibert}, \&
  {Scowcroft}}]{Neeley2019PLZ}
{Neeley}, J.~R., {Marengo}, M., {Freedman}, W.~L., {et~al.} 2019, \mnras, 490,
  4254

\bibitem[{{Nemec} {et~al.}(2013){Nemec}, {Cohen}, {Ripepi}, {Derekas},
  {Moskalik}, {Sesar}, {Chadid}, \& {Bruntt}}]{Nemec2013}
{Nemec}, J.~M., {Cohen}, J.~G., {Ripepi}, V., {et~al.} 2013, \apj, 773, 181

\bibitem[{{Netzel} \& {Smolec}(2022)}]{Netzel2022}
{Netzel}, H. \& {Smolec}, R. 2022, \mnras, 515, 3439

\bibitem[{{Netzel} {et~al.}(2015){Netzel}, {Smolec}, \&
  {Moskalik}}]{Netzel2015-02}
{Netzel}, H., {Smolec}, R., \& {Moskalik}, P. 2015, \mnras, 447, 1173

\bibitem[{P\'erez \& Granger(2007)}]{ipython}
P\'erez, F. \& Granger, B.~E. 2007, Computing in Science and Engineering, 9, 21

\bibitem[{{Petersen}(1986)}]{Petersen1986}
{Petersen}, J.~O. 1986, \aap, 170, 59

\bibitem[{{Pietrukowicz} {et~al.}(2015){Pietrukowicz}, {Koz{\l}owski},
  {Skowron}, {Soszy{\'n}ski}, {Udalski}, {Poleski}, {Wyrzykowski},
  {Szyma{\'n}ski}, {Pietrzy{\'n}ski}, {Ulaczyk}, {Mr{\'o}z}, {Skowron}, \&
  {Kubiak}}]{Pietrukowicz2015}
{Pietrukowicz}, P., {Koz{\l}owski}, S., {Skowron}, J., {et~al.} 2015, \apj,
  811, 113

\bibitem[{{Pietrzy{\'n}ski} {et~al.}(2019){Pietrzy{\'n}ski}, {Graczyk},
  {Gallenne}, {Gieren}, {Thompson}, {Pilecki}, {Karczmarek}, {G{\'o}rski},
  {Suchomska}, {Taormina}, {Zgirski}, {Wielg{\'o}rski}, {Ko{\l}aczkowski},
  {Konorski}, {Villanova}, {Nardetto}, {Kervella}, {Bresolin}, {Kudritzki},
  {Storm}, {Smolec}, \& {Narloch}}]{Pietrzynski2019}
{Pietrzy{\'n}ski}, G., {Graczyk}, D., {Gallenne}, A., {et~al.} 2019, \nat, 567,
  200

\bibitem[{{Pojmanski}(1997)}]{Pojmanski1997}
{Pojmanski}, G. 1997, \actaa, 47, 467

\bibitem[{{Prudil} {et~al.}(2019{\natexlab{a}}){Prudil}, {D{\'e}k{\'a}ny},
  {Catelan}, {Smolec}, {Grebel}, \& {Skarka}}]{Prudil2019OOspat}
{Prudil}, Z., {D{\'e}k{\'a}ny}, I., {Catelan}, M., {et~al.} 2019{\natexlab{a}},
  \mnras, 484, 4833

\bibitem[{{Prudil} {et~al.}(2019{\natexlab{b}}){Prudil}, {D{\'e}k{\'a}ny},
  {Grebel}, {Catelan}, {Skarka}, \& {Smolec}}]{Prudil2019Kin}
{Prudil}, Z., {D{\'e}k{\'a}ny}, I., {Grebel}, E.~K., {et~al.}
  2019{\natexlab{b}}, \mnras, 487, 3270

\bibitem[{{Sarajedini} {et~al.}(2006){Sarajedini}, {Barker}, {Geisler},
  {Harding}, \& {Schommer}}]{Sarajedini2006}
{Sarajedini}, A., {Barker}, M.~K., {Geisler}, D., {Harding}, P., \& {Schommer},
  R. 2006, \aj, 132, 1361

\bibitem[{{Sarajedini} {et~al.}(2009){Sarajedini}, {Mancone}, {Lauer},
  {Dressler}, {Freedman}, {Trager}, {Grillmair}, \& {Mighell}}]{Sarajedini2009}
{Sarajedini}, A., {Mancone}, C.~L., {Lauer}, T.~R., {et~al.} 2009, \aj, 138,
  184

\bibitem[{{Savino} {et~al.}(2022){Savino}, {Weisz}, {Skillman}, {Dolphin},
  {Kallivayalil}, {Wetzel}, {Anderson}, {Besla}, {Boylan-Kolchin}, {Bullock},
  {Cole}, {Collins}, {Cooper}, {Deason}, {Dotter}, {Fardal}, {Ferguson},
  {Fritz}, {Geha}, {Gilbert}, {Guhathakurta}, {Ibata}, {Irwin}, {Jeon},
  {Kirby}, {Lewis}, {Mackey}, {Majewski}, {Martin}, {McConnachie}, {Patel},
  {Rich}, {Simon}, {Sohn}, {Tollerud}, \& {van der Marel}}]{Savino2022}
{Savino}, A., {Weisz}, D.~R., {Skillman}, E.~D., {et~al.} 2022, arXiv e-prints,
  arXiv:2206.02801

\bibitem[{{Schlafly} \& {Finkbeiner}(2011)}]{Schlafly2011}
{Schlafly}, E.~F. \& {Finkbeiner}, D.~P. 2011, \apj, 737, 103

\bibitem[{{Schlegel} {et~al.}(1998){Schlegel}, {Finkbeiner}, \&
  {Davis}}]{Schlegel1998}
{Schlegel}, D.~J., {Finkbeiner}, D.~P., \& {Davis}, M. 1998, \apj, 500, 525

\bibitem[{{Shappee} {et~al.}(2014){Shappee}, {Prieto}, {Grupe}, {Kochanek},
  {Stanek}, {De Rosa}, {Mathur}, {Zu}, {Peterson}, {Pogge}, {Komossa}, {Im},
  {Jencson}, {Holoien}, {Basu}, {Beacom}, {Szczygie{\l}}, {Brimacombe},
  {Adams}, {Campillay}, {Choi}, {Contreras}, {Dietrich}, {Dubberley},
  {Elphick}, {Foale}, {Giustini}, {Gonzalez}, {Hawkins}, {Howell}, {Hsiao},
  {Koss}, {Leighly}, {Morrell}, {Mudd}, {Mullins}, {Nugent}, {Parrent},
  {Phillips}, {Pojmanski}, {Rosing}, {Ross}, {Sand}, {Terndrup}, {Valenti},
  {Walker}, \& {Yoon}}]{Shappee2014}
{Shappee}, B.~J., {Prieto}, J.~L., {Grupe}, D., {et~al.} 2014, \apj, 788, 48

\bibitem[{{Skarka} {et~al.}(2020){Skarka}, {Prudil}, \& {Jurcsik}}]{Skarka2020}
{Skarka}, M., {Prudil}, Z., \& {Jurcsik}, J. 2020, \mnras, 494, 1237

\bibitem[{{Skowron} {et~al.}(2016){Skowron}, {Soszy{\'n}ski}, {Udalski},
  {Szyma{\'n}ski}, {Pietrukowicz}, {Skowron}, {Poleski}, {Wyrzykowski},
  {Ulaczyk}, {Koz{\l}owski}, {Mr{\'o}z}, \& {Pawlak}}]{Skowron2016}
{Skowron}, D.~M., {Soszy{\'n}ski}, I., {Udalski}, A., {et~al.} 2016, \actaa,
  66, 269

\bibitem[{{Skrutskie} {et~al.}(2006){Skrutskie}, {Cutri}, {Stiening},
  {Weinberg}, {Schneider}, {Carpenter}, {Beichman}, {Capps}, {Chester},
  {Elias}, {Huchra}, {Liebert}, {Lonsdale}, {Monet}, {Price}, {Seitzer},
  {Jarrett}, {Kirkpatrick}, {Gizis}, {Howard}, {Evans}, {Fowler}, {Fullmer},
  {Hurt}, {Light}, {Kopan}, {Marsh}, {McCallon}, {Tam}, {Van Dyk}, \&
  {Wheelock}}]{Skrutskie2006}
{Skrutskie}, M.~F., {Cutri}, R.~M., {Stiening}, R., {et~al.} 2006, \aj, 131,
  1163

\bibitem[{{Smolec} {et~al.}(2015){Smolec}, {Soszy{\'n}ski}, {Udalski},
  {Szyma{\'n}ski}, {Pietrukowicz}, {Skowron}, {Koz{\l}owski}, {Poleski},
  {Skowron}, {Pietrzy{\'n}ski}, {Wyrzykowski}, {Ulaczyk}, \&
  {Mr{\'o}z}}]{Smolec2015-BL}
{Smolec}, R., {Soszy{\'n}ski}, I., {Udalski}, A., {et~al.} 2015, \mnras, 447,
  3756

\bibitem[{{Sneden} {et~al.}(2017){Sneden}, {Preston}, {Chadid}, \&
  {Adam{\'o}w}}]{Sneden2017}
{Sneden}, C., {Preston}, G.~W., {Chadid}, M., \& {Adam{\'o}w}, M. 2017, \apj,
  848, 68

\bibitem[{{Sollima} {et~al.}(2008){Sollima}, {Cacciari}, {Arkharov},
  {Larionov}, {Gorshanov}, {Efimova}, \& {Piersimoni}}]{Sollima2008}
{Sollima}, A., {Cacciari}, C., {Arkharov}, A.~A.~H., {et~al.} 2008, \mnras,
  384, 1583

\bibitem[{{Sollima} {et~al.}(2006){Sollima}, {Cacciari}, \&
  {Valenti}}]{Sollima2006}
{Sollima}, A., {Cacciari}, C., \& {Valenti}, E. 2006, \mnras, 372, 1675

\bibitem[{{Soszy{\'n}ski} {et~al.}(2016){Soszy{\'n}ski}, {Udalski},
  {Szyma{\'n}ski}, {Wyrzykowski}, {Ulaczyk}, {Poleski}, {Pietrukowicz},
  {Koz{\l}owski}, {Skowron}, {Skowron}, {Mr{\'o}z}, \&
  {Pawlak}}]{Soszynski2016LMCSMC}
{Soszy{\'n}ski}, I., {Udalski}, A., {Szyma{\'n}ski}, M.~K., {et~al.} 2016,
  \actaa, 66, 131

\bibitem[{{Stetson} {et~al.}(2014){Stetson}, {Braga}, {Dall'Ora}, {Bono},
  {Buonanno}, {Ferraro}, {Iannicola}, {Marengo}, \& {Neeley}}]{Stetson2014}
{Stetson}, P.~B., {Braga}, V.~F., {Dall'Ora}, M., {et~al.} 2014, \pasp, 126,
  521

\bibitem[{{Szab{\'o}} {et~al.}(2014){Szab{\'o}}, {Benk{\H{o}}}, {Papar{\'o}},
  {Chapellier}, {Poretti}, {Baglin}, {Weiss}, {Kolenberg}, {Guggenberger}, \&
  {Le Borgne}}]{LeBorgne2014}
{Szab{\'o}}, R., {Benk{\H{o}}}, J.~M., {Papar{\'o}}, M., {et~al.} 2014, \aap,
  570, A100

\bibitem[{{Szab{\'o}} {et~al.}(2010){Szab{\'o}}, {Koll{\'a}th}, {Moln{\'a}r},
  {Kolenberg}, {Kurtz}, {Bryson}, {Benk{\H o}}, {Christensen-Dalsgaard},
  {Kjeldsen}, {Borucki}, {Koch}, {Twicken}, {Chadid}, {di Criscienzo}, {Jeon},
  {Moskalik}, {Nemec}, \& {Nuspl}}]{Szabo2010}
{Szab{\'o}}, R., {Koll{\'a}th}, Z., {Moln{\'a}r}, L., {et~al.} 2010, \mnras,
  409, 1244

\bibitem[{{Szczygie{\l}} {et~al.}(2009){Szczygie{\l}}, {Pojma{\'n}ski}, \&
  {Pilecki}}]{Szczygiel2009}
{Szczygie{\l}}, D.~M., {Pojma{\'n}ski}, G., \& {Pilecki}, B. 2009, \actaa, 59,
  137

\bibitem[{{Tatton} {et~al.}(2021){Tatton}, {van Loon}, {Cioni}, {Bekki},
  {Bell}, {Choudhury}, {de Grijs}, {Groenewegen}, {Ivanov}, {Marconi},
  {Oliveira}, {Ripepi}, {Rubele}, {Subramanian}, \& {Sun}}]{Tatton2021}
{Tatton}, B.~L., {van Loon}, J.~T., {Cioni}, M. R.~L., {et~al.} 2021, \mnras,
  504, 2983

\bibitem[{{Udalski} {et~al.}(2015){Udalski}, {Szyma{\'n}ski}, \&
  {Szyma{\'n}ski}}]{Udalski2015}
{Udalski}, A., {Szyma{\'n}ski}, M.~K., \& {Szyma{\'n}ski}, G. 2015, \actaa, 65,
  1

\bibitem[{{van Leeuwen}(2007)}]{Leeuwen2007cat}
{van Leeuwen}, F. 2007, \aap, 474, 653

\bibitem[{{Vasiliev} \& {Baumgardt}(2021)}]{Vasiliev2021}
{Vasiliev}, E. \& {Baumgardt}, H. 2021, \mnras, 505, 5978

\bibitem[{Virtanen {et~al.}(2020)Virtanen, Gommers, Oliphant, Haberland, Reddy,
  Cournapeau, Burovski, Peterson, Weckesser, Bright, {van der Walt}, Brett,
  Wilson, Millman, Mayorov, Nelson, Jones, Kern, Larson, Carey, Polat, Feng,
  Moore, {VanderPlas}, Laxalde, Perktold, Cimrman, Henriksen, Quintero, Harris,
  Archibald, Ribeiro, Pedregosa, {van Mulbregt}, \& {SciPy 1.0
  Contributors}}]{scipy}
Virtanen, P., Gommers, R., Oliphant, T.~E., {et~al.} 2020, Nature Methods, 17,
  261

\bibitem[{{Watson} {et~al.}(2006){Watson}, {Henden}, \&
  {Price}}]{Watson2006VSX}
{Watson}, C.~L., {Henden}, A.~A., \& {Price}, A. 2006, Society for Astronomical
  Sciences Annual Symposium, 25, 47

\bibitem[{{Zinn} \& {West}(1984)}]{Zinn1984}
{Zinn}, R. \& {West}, M.~J. 1984, \apjs, 55, 45

\end{thebibliography}

\begin{appendix} 
\section{Additional figures} \label{sec:AddFigures}

\begin{figure}
\includegraphics[width=\columnwidth]{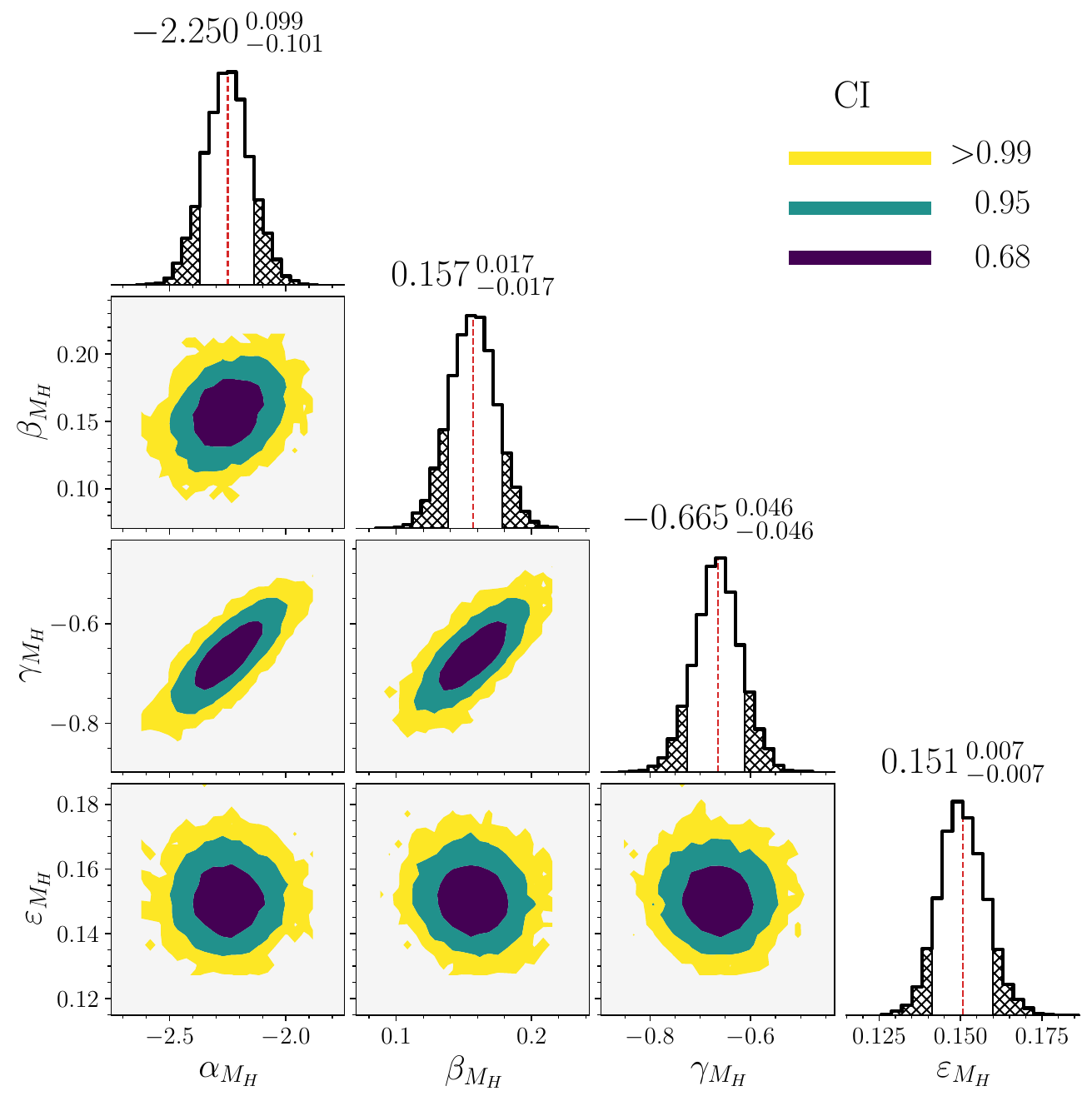}
\caption{Same as Fig.~\ref{fig:CornerKs} but for the $H$-band.}
\label{fig:CornerH}
\end{figure}

\begin{figure}
\includegraphics[width=\columnwidth]{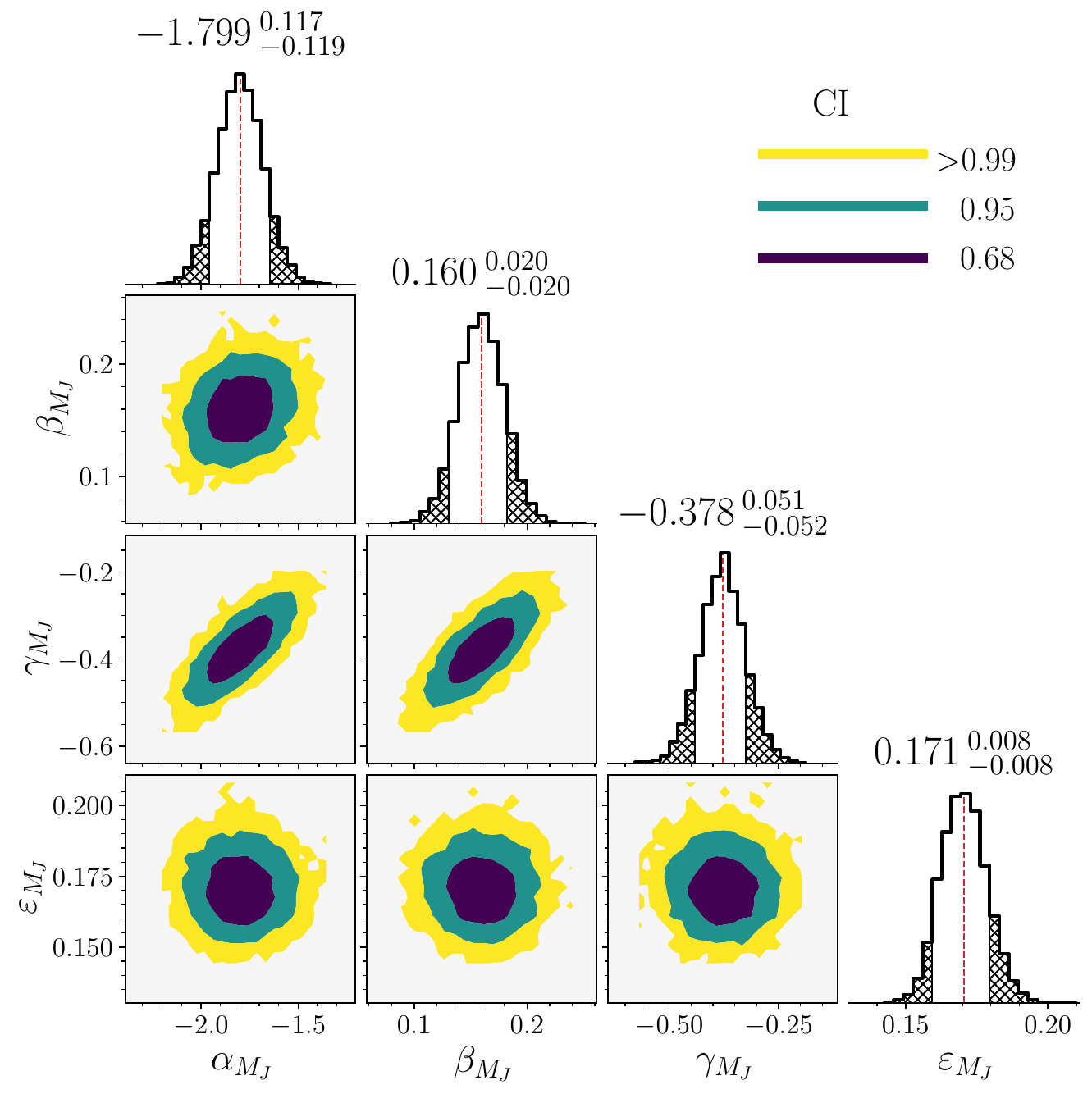}
\caption{Same as Fig.~\ref{fig:CornerKs} but for the $J$-band.}
\label{fig:CornerJ}
\end{figure}

\begin{figure}
\includegraphics[width=\columnwidth]{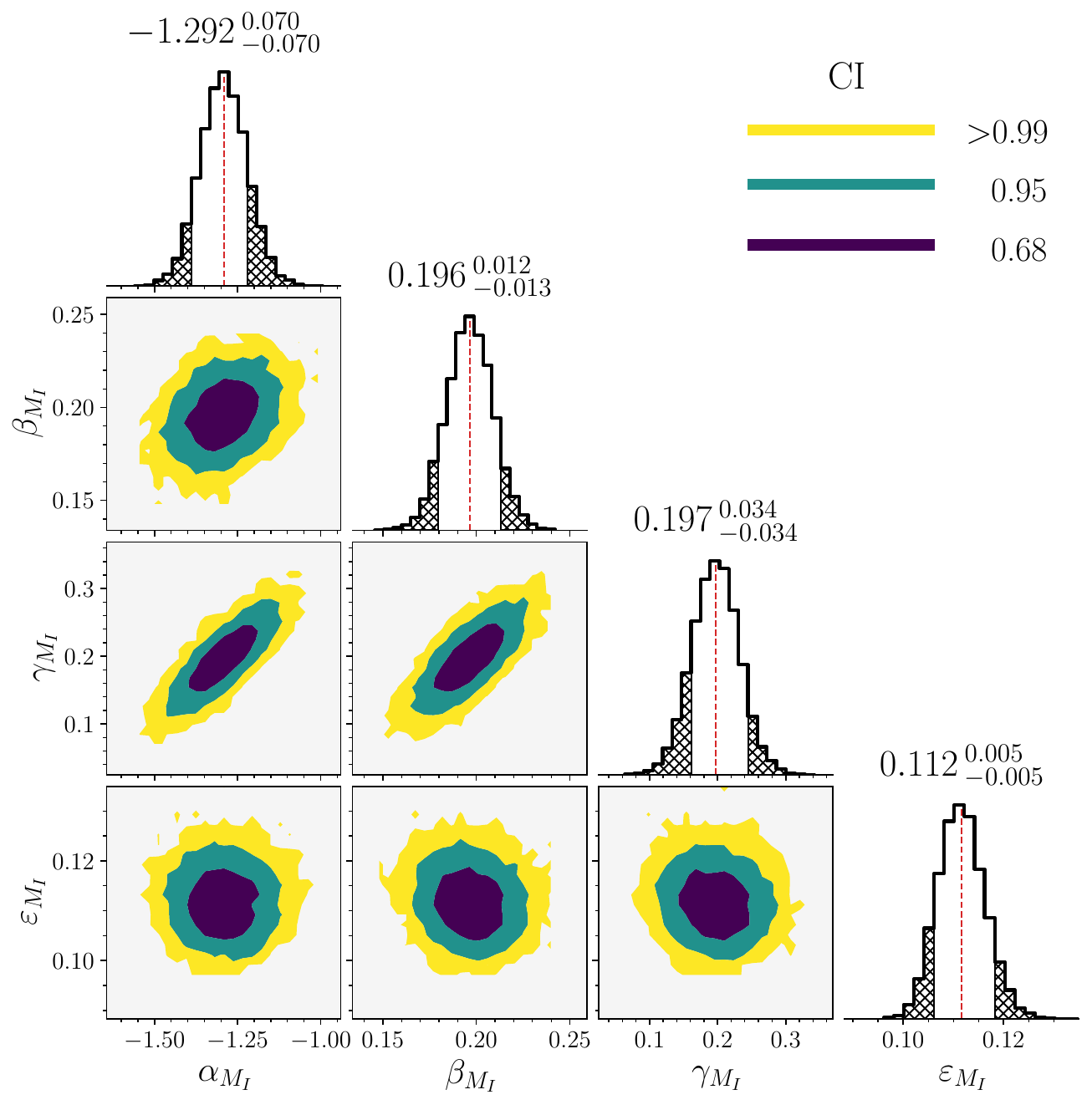}
\caption{Same as Fig.~\ref{fig:CornerKs} but for the $I$-band.}
\label{fig:CornerI}
\end{figure}

\begin{figure}
\includegraphics[width=\columnwidth]{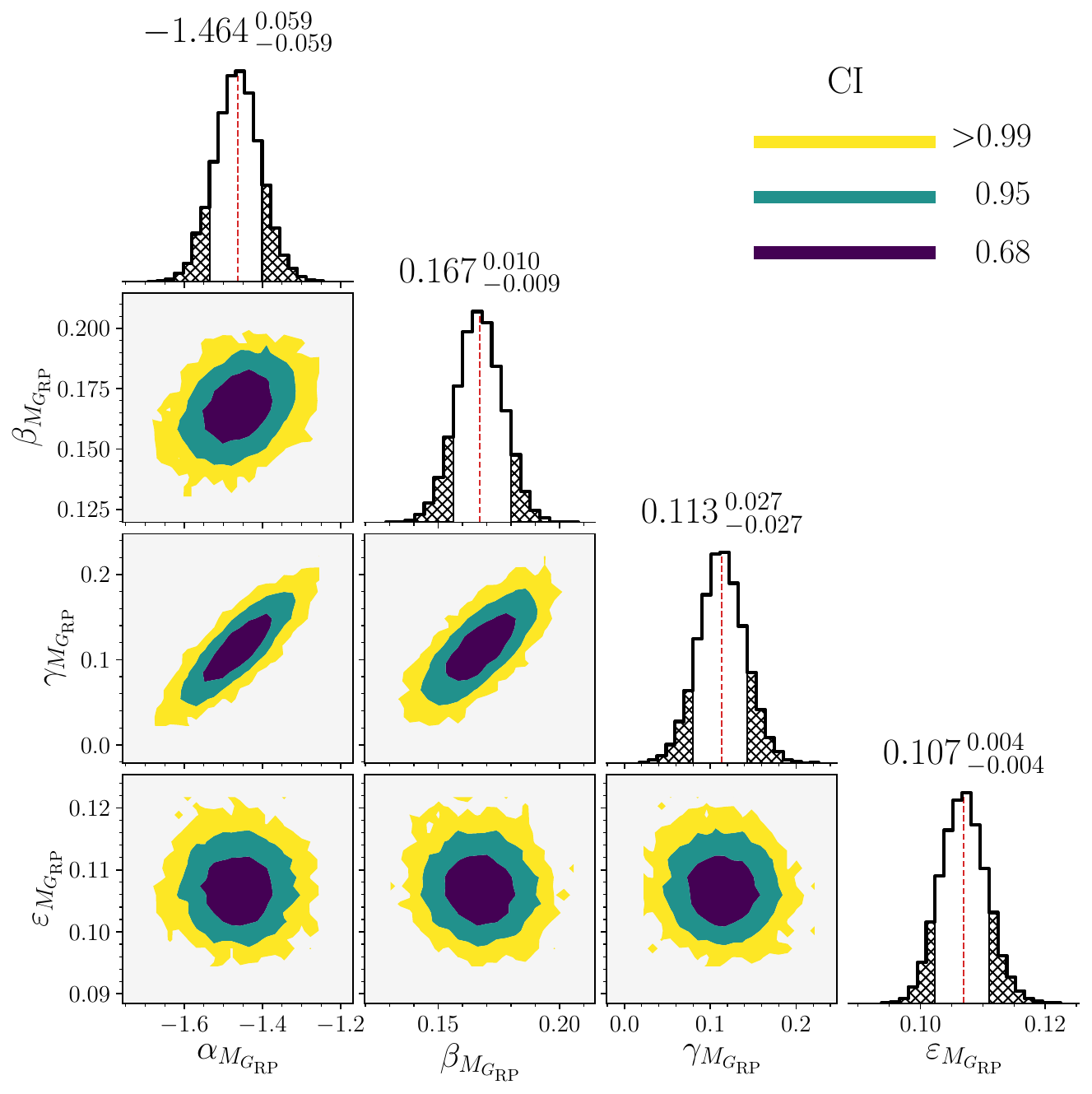}
\caption{Same as Fig.~\ref{fig:CornerKs} but for the $G_{\rm RP}$-band.}
\label{fig:CornerGrp}
\end{figure}

\begin{figure}
\includegraphics[width=\columnwidth]{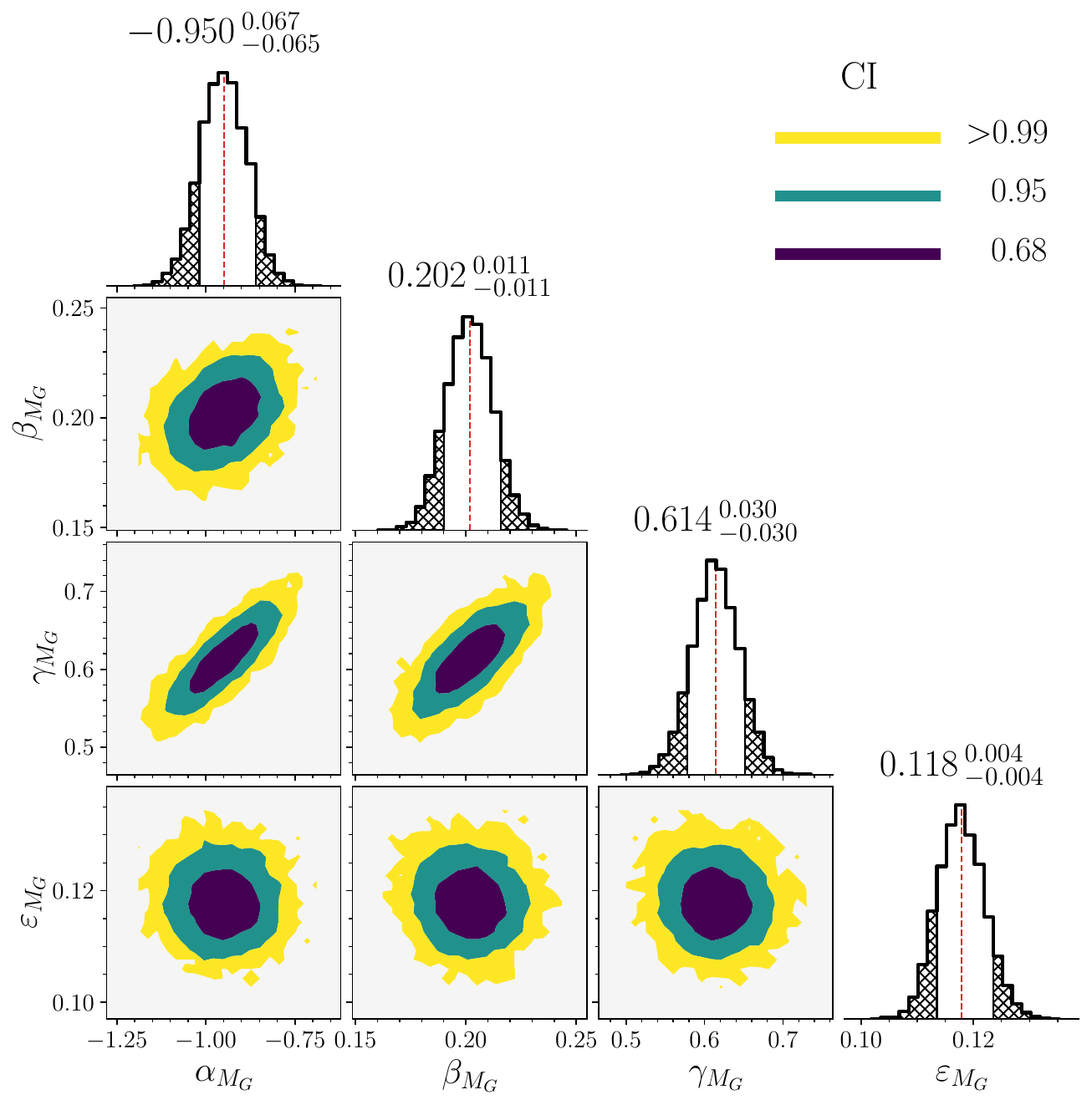}
\caption{Same as Fig.~\ref{fig:CornerKs} but for the $G$-band.}
\label{fig:CornerG}
\end{figure}

\begin{figure}
\includegraphics[width=\columnwidth]{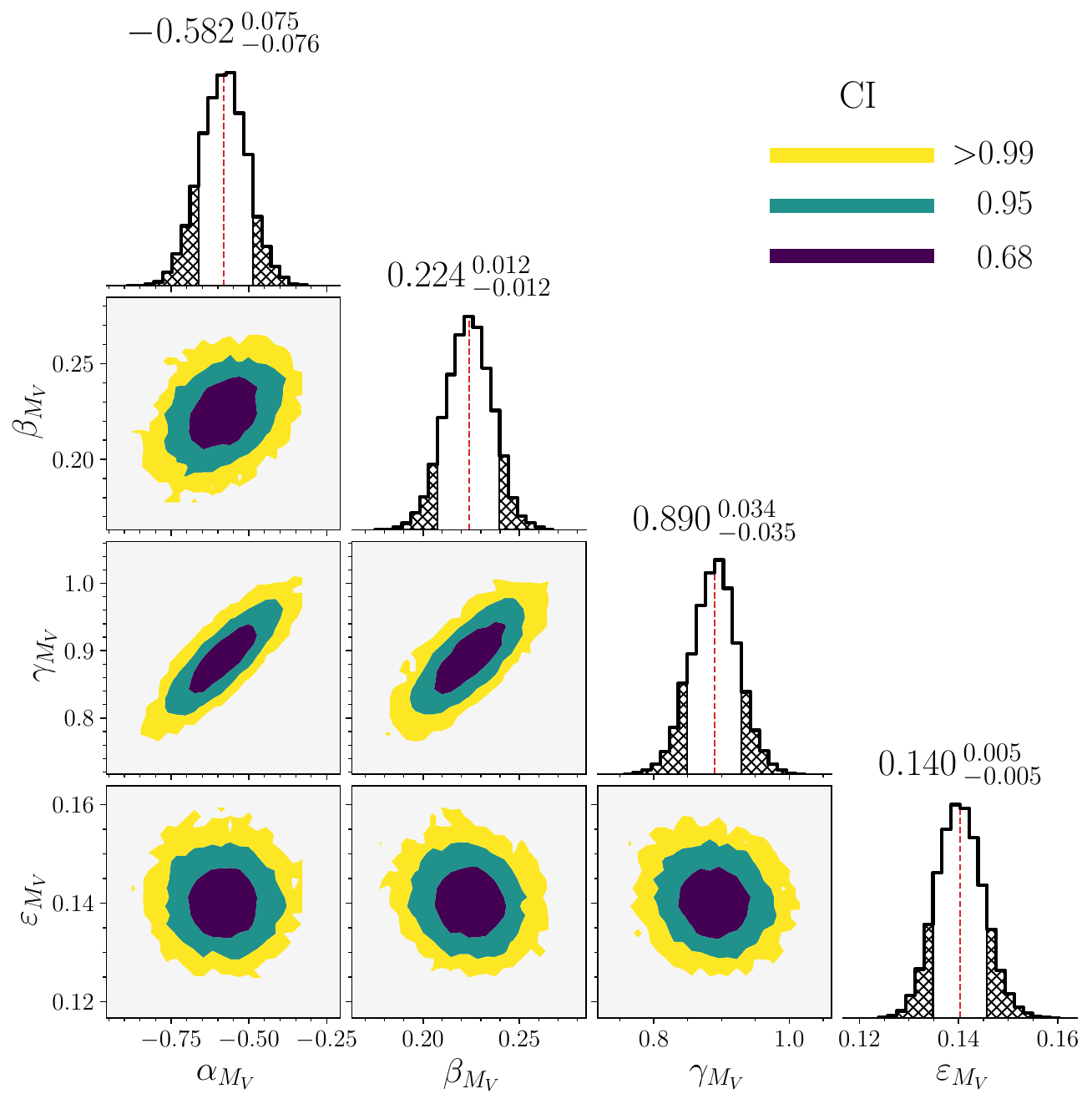}
\caption{Same as Fig.~\ref{fig:CornerKs} but for the $V$-band.}
\label{fig:CornerV}
\end{figure}

\begin{figure}
\includegraphics[width=\columnwidth]{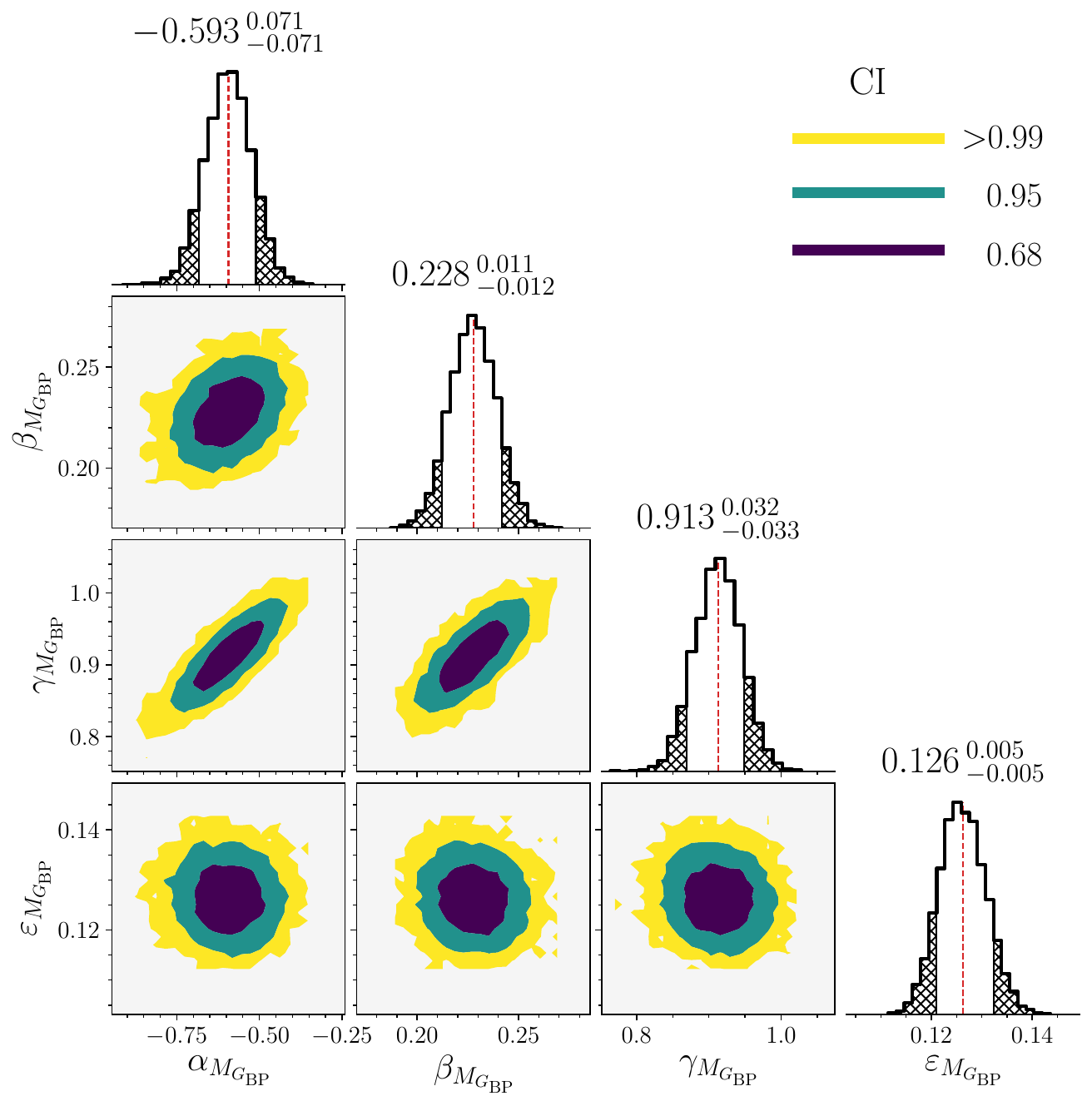}
\caption{Same as Fig.~\ref{fig:CornerKs} but for the $G_{\rm BP}$-band.}
\label{fig:CornerGbp}
\end{figure}

\begin{figure}
\includegraphics[width=\columnwidth]{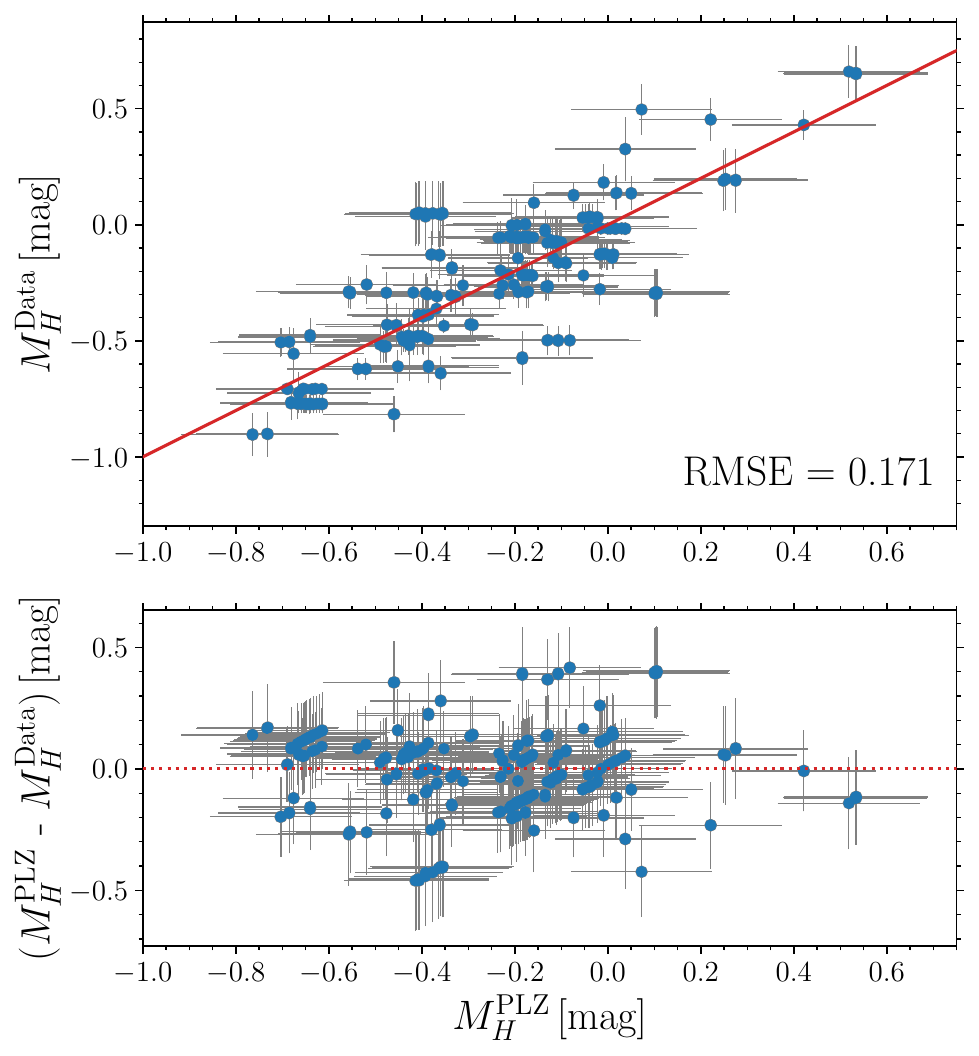}
\caption{Same as Fig.~\ref{fig:PMZ_comK} but for the $J$-band.}
\label{fig:PMZ_comH}
\end{figure}

\begin{figure}
\includegraphics[width=\columnwidth]{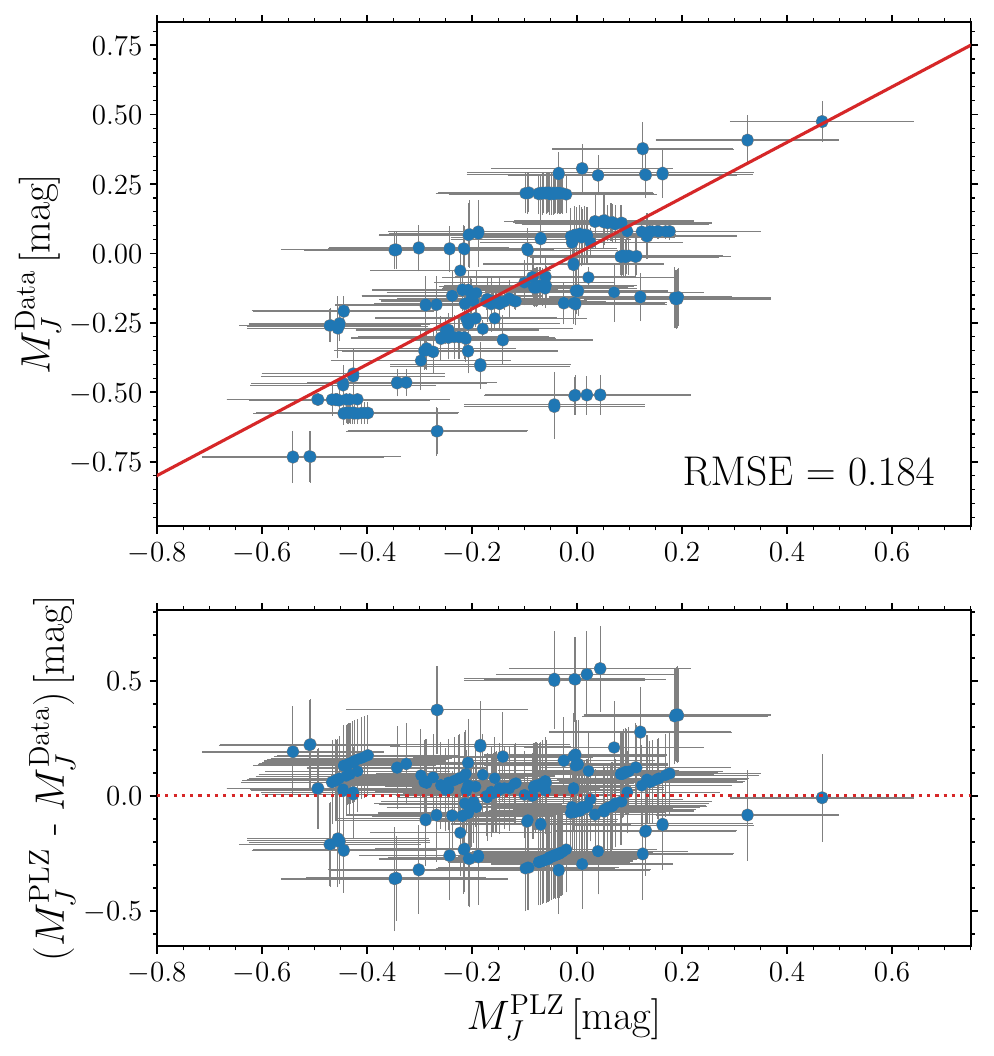}
\caption{Same as Fig.~\ref{fig:PMZ_comK} but for the $J$-band.}
\label{fig:PMZ_comJ}
\end{figure}

\begin{figure}
\includegraphics[width=\columnwidth]{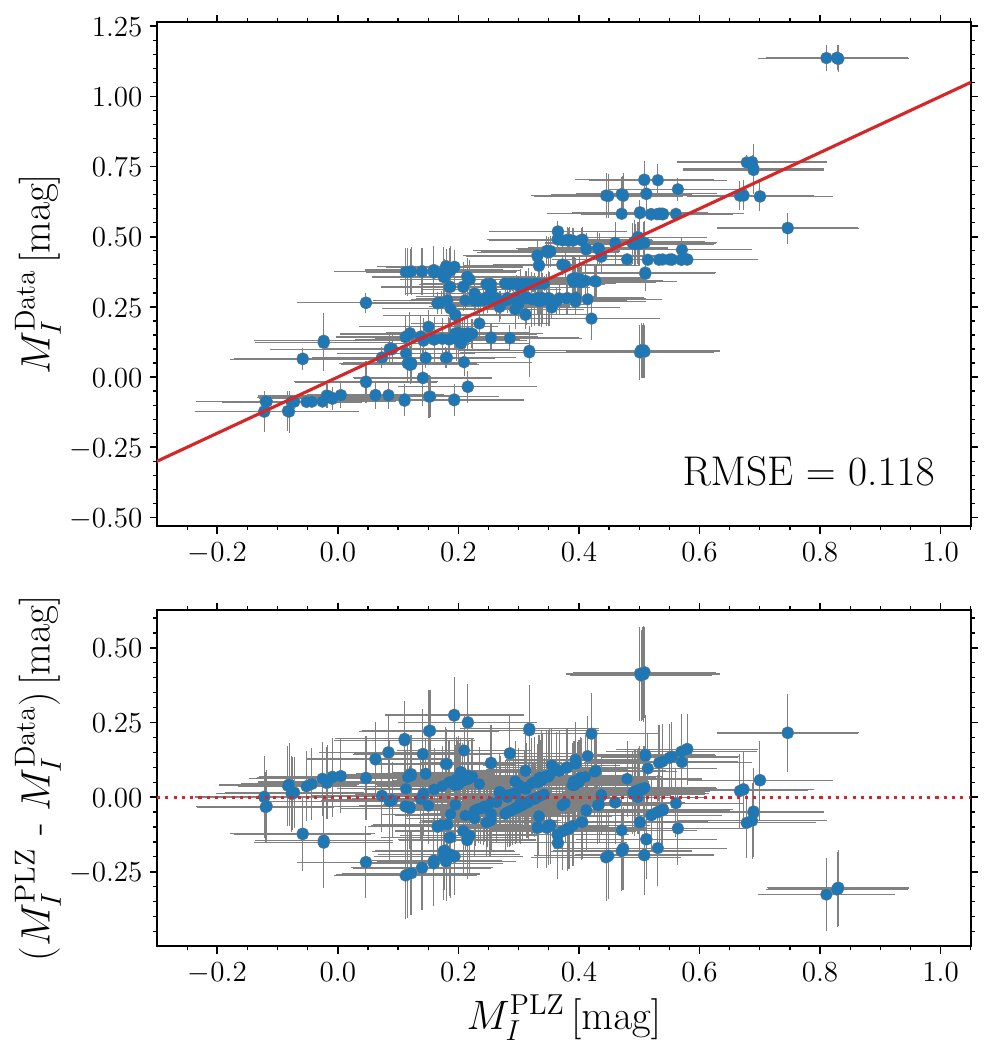}
\caption{Same as Fig.~\ref{fig:PMZ_comK} but for the $I$-band.}
\label{fig:PMZ_comI}
\end{figure}

\begin{figure}
\includegraphics[width=\columnwidth]{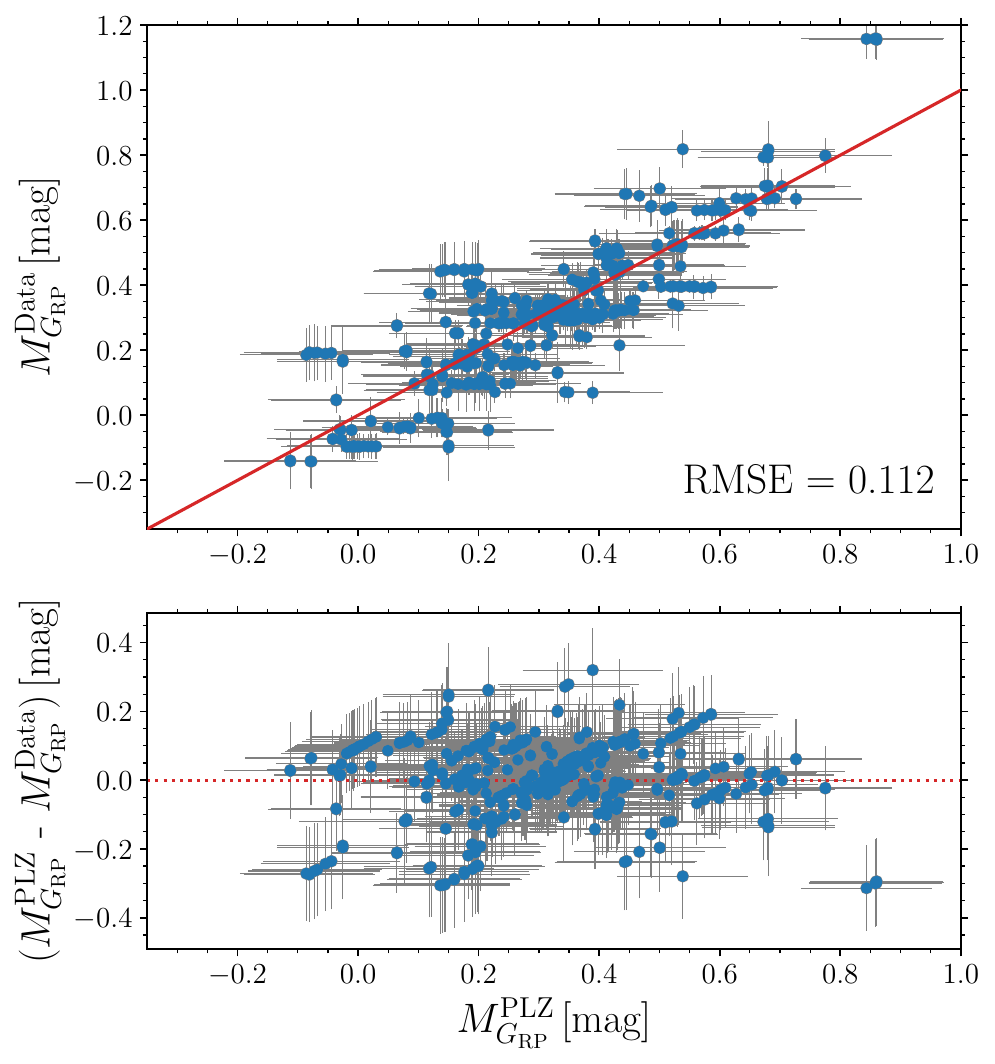}
\caption{Same as Fig.~\ref{fig:PMZ_comK} but for the $G_{\rm RP}$-band.}
\label{fig:PMZ_comGrp}
\end{figure}

\begin{figure}
\includegraphics[width=\columnwidth]{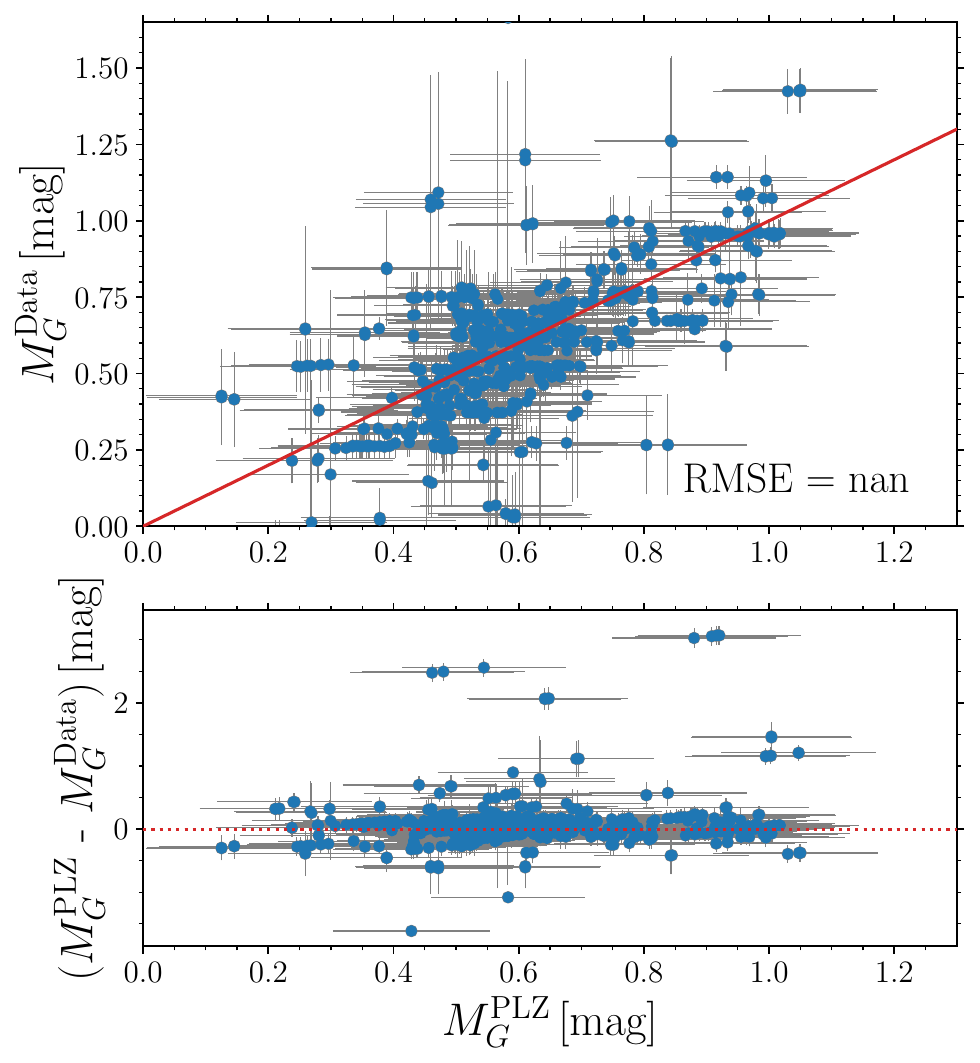}
\caption{Same as Fig.~\ref{fig:PMZ_comK} but for the $G$-band.}
\label{fig:PMZ_comG}
\end{figure}

\begin{figure}
\includegraphics[width=\columnwidth]{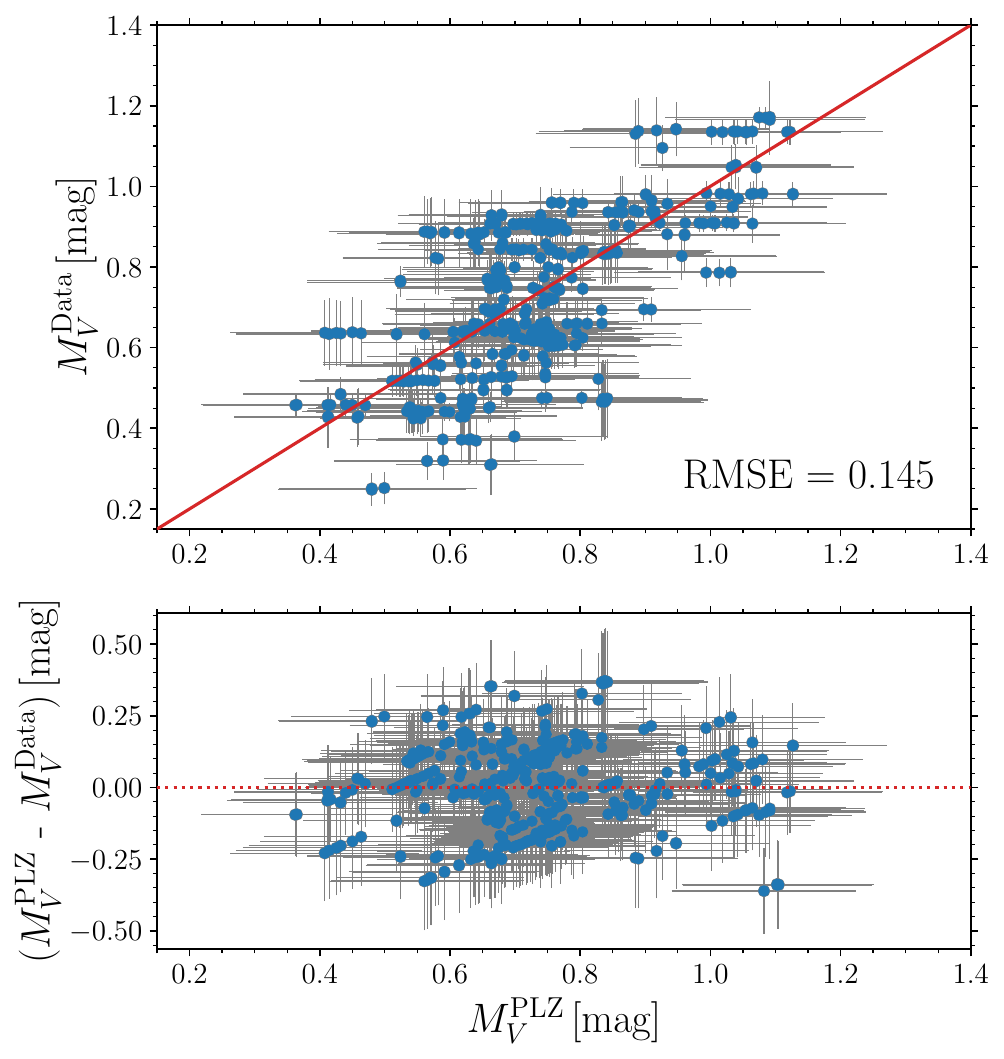}
\caption{Same as Fig.~\ref{fig:PMZ_comK} but for the $V$-band.}
\label{fig:PMZ_comV}
\end{figure}

\begin{figure}
\includegraphics[width=\columnwidth]{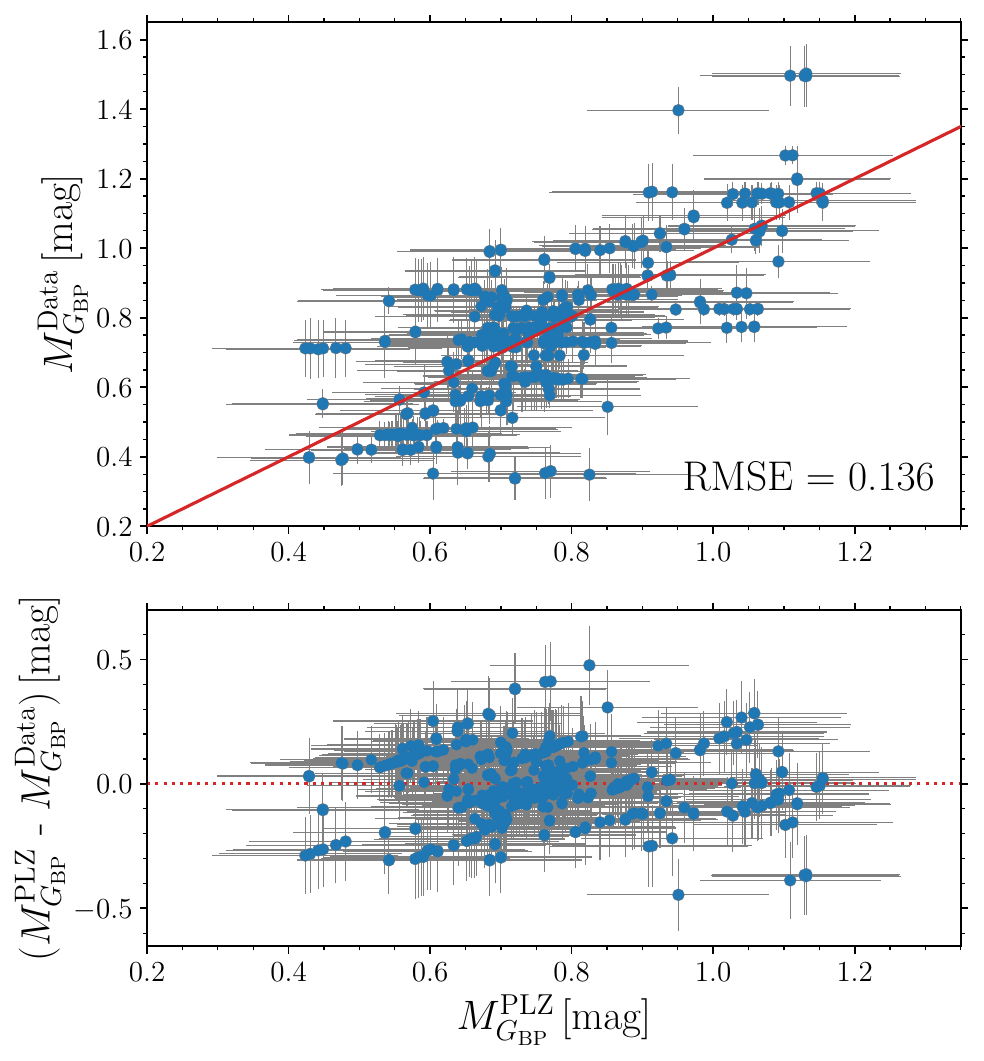}
\caption{Same as Fig.~\ref{fig:PMZ_comK} but for the $G_{\rm BP}$-band.}
\label{fig:PMZ_comGbp}
\end{figure}

\end{appendix}

\end{document}